\documentclass[aps,pra,reprint,superscriptaddress,amsmath,amssymb,floatfix]{revtex4-2}

\usepackage{graphicx}
\usepackage{bm}
\usepackage{hyperref}

\usepackage{lmodern}
\usepackage[cm]{sfmath}

\usepackage[font={sf, small}, justification=RaggedRight,labelfont=bf]{caption}

\begin{document}

\title{Single chip photonic deep neural network with accelerated training}
\author{Saumil Bandyopadhyay}
\email{saumilb@mit.edu}
\affiliation{Research Laboratory of Electronics, MIT, Cambridge, MA 02139, USA}

\author{Alexander Sludds}
\affiliation{Research Laboratory of Electronics, MIT, Cambridge, MA 02139, USA}

\author{Stefan Krastanov}
\affiliation{Research Laboratory of Electronics, MIT, Cambridge, MA 02139, USA}

\author{Ryan Hamerly}
\affiliation{Research Laboratory of Electronics, MIT, Cambridge, MA 02139, USA}
\affiliation{NTT Research Inc., PHI Laboratories, 940 Stewart Drive, Sunnyvale, CA 94085, USA}

\author{Nicholas Harris}
\affiliation{Research Laboratory of Electronics, MIT, Cambridge, MA 02139, USA}

\author{Darius Bunandar}
\affiliation{Research Laboratory of Electronics, MIT, Cambridge, MA 02139, USA}

\author{Matthew Streshinsky}
\affiliation{Nokia Corporation, New York, NY, 10016, USA}

\author{Michael Hochberg}
\affiliation{Luminous Computing Inc., Mountain View, CA, 94041, USA}

\author{Dirk Englund}
\email{englund@mit.edu}
\affiliation{Research Laboratory of Electronics, MIT, Cambridge, MA 02139, USA}

\begin{abstract}
\noindent As deep neural networks (DNNs) revolutionize machine learning \cite{NIPS2012_c399862d, 7780459, NEURIPS2020_1457c0d6, mirhoseini_graph_2021, vinyals_grandmaster_2019, silver_mastering_2017}, energy consumption and throughput are emerging as fundamental limitations of CMOS electronics. This has motivated a search for new hardware architectures optimized for artificial intelligence, such as electronic systolic arrays \cite{Jouppi2017}, memristor crossbar arrays \cite{xia_memristive_2019}, and optical accelerators.
Optical systems can perform linear matrix operations at exceptionally high rate and efficiency \cite{wetzstein_inference_2020}, motivating recent demonstrations of low latency linear algebra \cite{shen_deep_2017, bernstein_single-shot_2022, xu_11_2021, feldmann_parallel_2021, ashtiani_-chip_2022} and optical energy consumption \cite{wang_optical_2022, sludds_delocalized_2022} below a  photon per multiply-accumulate operation. However, demonstrating systems that co-integrate both linear and nonlinear processing units in a single chip remains a central challenge. Here we introduce such a system in a scalable photonic integrated circuit (PIC), enabled by several key advances: (i) high-bandwidth and low-power programmable nonlinear optical function units (NOFUs); (ii) coherent matrix multiplication units (CMXUs); and (iii) \textit{in situ} training with optical acceleration. We experimentally demonstrate this fully-integrated coherent optical neural network (FICONN) architecture for a 3-layer DNN comprising 12 NOFUs and three CMXUs operating in the telecom C-band. Using \textit{in situ} training on a vowel classification task, the FICONN achieves 92.7\% accuracy on a test set, which is identical to the accuracy obtained on a digital computer with the same number of weights. This work lends experimental evidence to theoretical proposals for \textit{in situ} training, unlocking orders of magnitude improvements in the throughput of training data. Moreover, the FICONN opens the path to inference at nanosecond latency and femtojoule per operation energy efficiency.
\end{abstract}

\maketitle

State-of-the-art artificial intelligence algorithms, including deep neural networks (DNNs), require vast amounts of computation that is mostly dominated by linear algebra. The sheer amount of computation required, together with thermal cooling limits, has motivated the development of new hardware in which energy efficiency is a key design parameter. This includes ``output-stationary'' optical hardware \cite{hamerly_large-scale_2019, sludds_delocalized_2022}, which compute matrix-matrix products by integrating optical signals over multiple time steps, and ``weight-stationary'' architectures \cite{shen_deep_2017, xu_11_2021, wang_optical_2022,  feldmann_parallel_2021, tait_broadcast_2014, ashtiani_-chip_2022,bernstein_single-shot_2022} that compute a single matrix-vector product per clock cycle. 

Weight-stationary architectures in particular are well-suited for applications that require processing data natively in the optical domain and with ultra-low latency. For example, self-driving cars require making split-second decisions by processing and learning from sensor readings acquired by LiDAR systems \cite{liu_efficient_2021}; scientific research in astronomy \cite{messick_analysis_2017, huerta_enabling_2019} and particle physics \cite{coelho_automatic_2021} requires rapid analysis of weak signals; and recently introduced ``smart'' optical transceivers rely on machine learning to receive, process, and route data at line rates exceeding hundreds of gigabits per second \cite{zibar_machine_2016, sludds_delocalized_2022}. These applications present an opportunity for real-time inference and training directly on optical signals, preserving phase information and avoiding the need for electrical-to-optical conversions. 

Here we report the first demonstration of a fully-integrated, coherent optical deep neural network that performs inference and \textit{in situ} training. This is made possible by three key advances spanning devices to system hardware to algorithms:
\begin{enumerate} 
\item \textit{Coherent programmable optical nonlinearities}: An outstanding challenge for optical DNNs is realizing fast, energy-efficient nonlinearities that can be integrated into photonic circuits. Electro-optical activation functions have previously been realized in silicon photonics, but required off-chip, high-power electronic amplifiers \cite{williamson_reprogrammable_2020, ashtiani_-chip_2022} to generate a sufficient nonlinear response. Here we design a nonlinear optical function unit (NOFU) for DNNs that is fabricated in a commercial foundry process, does not require an amplifier, and implements a reconfigurable coherent nonlinear operation on the optical field.
\item \textit{Coherent matrix multiplication units}: Computing linear algebra coherently introduces the opportunity to process optical signals, which contain information in both amplitude and phase, directly within DNNs while bypassing slow optical-to-electronic conversions. We realize the DNN's linear transformations with a coherent matrix multiplication unit (CMXU) that computes matrix-vector products through passive interference in a Mach-Zehnder interferometer (MZI) mesh \cite{miller_self-configuring_2013}. While previous demonstrations have used individual MZI meshes to implement the linear layer of a DNN \cite{shen_deep_2017, zhang_optical_2021}, here we integrate multiple CMXUs to realize a coherent optical DNN on a single chip. 
\item \textit{In situ training of coherent optical DNNs}: The efficient training of model parameters is a central challenge in machine learning. 
A critical bottleneck for model training is forward inference, as it requires many evaluations of the model on a large training set to optimize weight parameters. \textit{In situ} training on photonic hardware can take advantage of  near-instantaneous DNN inference,  lowering the latency and power consumption of model training. Moreover, learning weights in real time can benefit applications that natively process optical data, such as LiDAR systems \cite{liu_efficient_2021},  optical transceivers~\cite{zibar_machine_2016}, and federated learning for edge devices~\cite{konecny_federated_2015, sludds_delocalized_2022}.\\ 
\\
While prior work has shown training of individual linear layers of photonic neural networks \textit{in situ}
\cite{pai2022, zhang_efficient_2021}, finding efficient algorithms for training all-photonic DNNs, which include on-chip optical nonlinearities, remains a major obstacle. 
Here we report, to the best of our knowledge, the first demonstration of \textit{in situ} training of a fully-integrated photonic DNN. We demonstrate efficient, optically-accelerated training of the hardware to perform a vowel classification task with 92.7\% accuracy, which is the same as the accuracy obtained on a digital computer. Our approach, which computes derivatives directly on hardware, generalizes beyond the system presented here to the many other photonic architectures for DNNs being currently studied.
\end{enumerate}
Our system, comprising 132 individually tunable model parameters on-chip, is the largest integrated photonic DNN demonstrated to date in number of weights and computes matrix operations and nonlinear activation functions coherently on optical fields. We combined all subsystems for a coherent optical DNN into a photonic integrated circuit (PIC), fabricated in a commercial silicon photonics process, that incorporates both programmable linear and nonlinear transformations onto a single 6 $\times$ 5.7 mm$^2$ chip.

\begin{figure*}
    \includegraphics[width=7in]{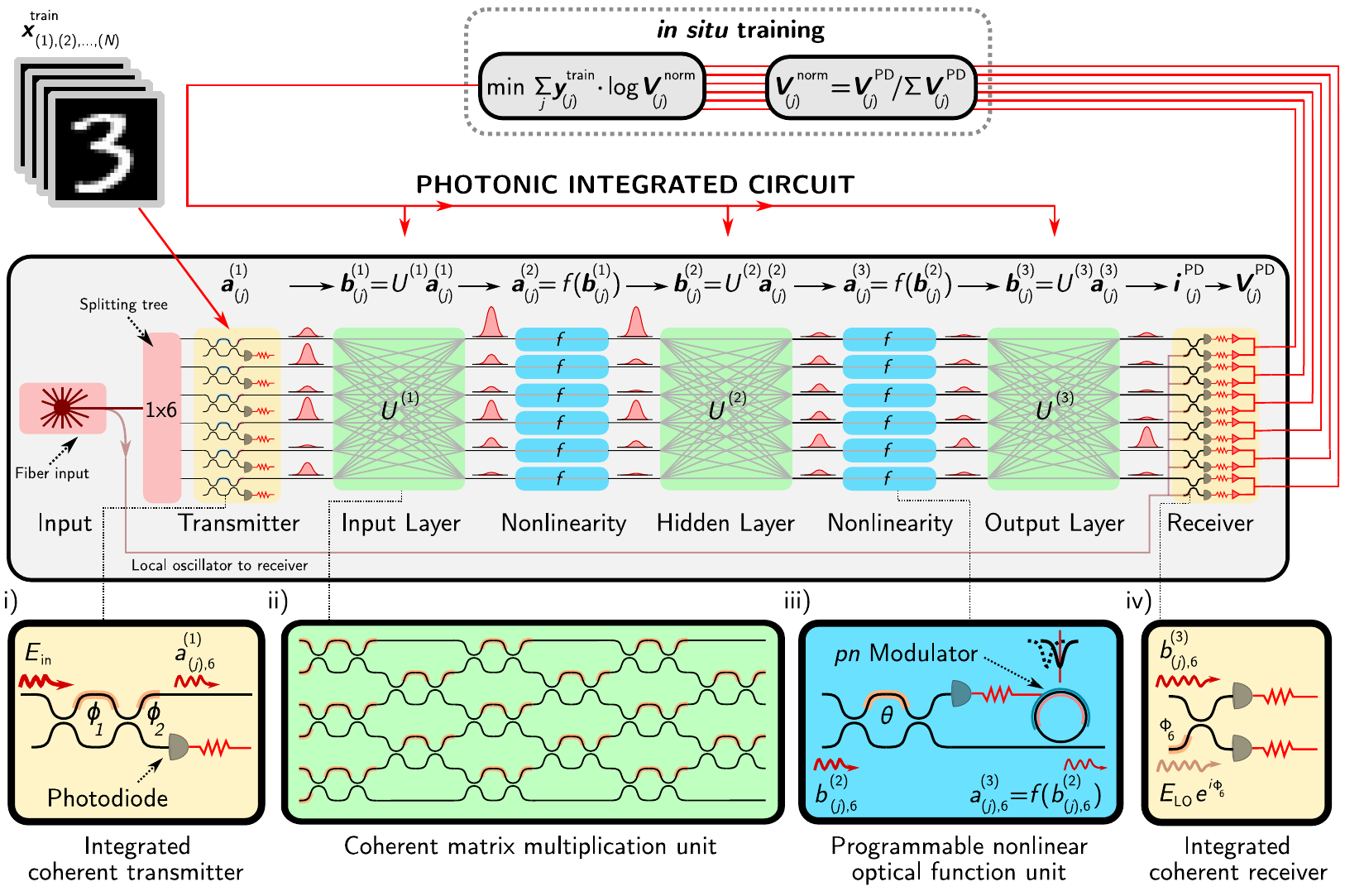}
    \caption{Architecture of the fully-integrated coherent optical neural network (FICONN). Inference is conducted entirely in the optical domain, without readout or amplification between layers. Light is fiber coupled into a single input on the chip and fanned out to the six channels of the transmitter \textbf{(i)}. Each channel encodes the amplitude and phase of one element of the input $\mathbf{x}_{(j)}$ into the optical field $\mathbf{a}^{(1)}_{(j)}$ with a Mach-Zehnder modulator and an external phase shifter. The coherent matrix multiplication unit \textbf{(ii)}, consisting of a Mach-Zehnder interferometer mesh, implements the linear transformation $\mathbf{b}_{(j)}^{(n)} = U^{(n)} \mathbf{a}_{(j)}^{(n)}$. Programmable nonlinear optical function units \textbf{(iii)} realize activation functions $\mathbf{a}_{(j)}^{(n+1)} = f(\mathbf{b}_{(j)}^{(n)})$ by tapping off part of the signal to a photodiode, which drives a cavity off-resonance by injecting carriers into the waveguide. An integrated coherent receiver \textbf{(iv)} reads out the DNN output by homodyning the output field with a local oscillator. Transimpedance amplifiers convert the output photocurrents to voltages, which are  digitized and normalized to produce a quasi-probability distribution for a classification task.
    During \textit{in situ} training, the model parameters $\mathbf{\Theta}$ are recurrently optimized to minimize the categorical cross-entropy  over the training set $\mathbf{x}^{\mathrm{train}}_{(1),(2),...(N)}$.}
    \label{architecture}
\end{figure*}

\section*{Architecture}

Figure \ref{architecture} shows how this PIC architecture allows us to realize a fully integrated  coherent optical neural network (FICONN) through the following stages: (i) the \textit{transmitter} (TX) maps input vectors $\mathbf{x}_{(j)}$ to an optical field vector $\mathbf{a}_{(j)}^{(1)}$ by splitting an input laser field into MZI modulators $m=1,2,...,6$, each of which encode one element of $\mathbf{x}_{(j)}$ into the amplitude $A_m$ and phase $\phi_m$ of the transverse electric field component $a_{(j),m}^{(1)} = A_m e^{i \omega t + \phi_m}$;  (ii) the \textit{coherent matrix multiplication unit} (CMXU), consisting of a MZI mesh \cite{bogaerts_programmable_2020,shen_deep_2017,zhang_optical_2021}, transforms $\mathbf{a}^{(1)} \rightarrow \mathbf{b}^{(1)} = U^{(1)} \mathbf{a}^{(1)}$ through passive optical interference; and (iii) the \textit{programmable nonlinear optical function unit} (NOFU) applies the activation function to yield the input to the next layer, $\mathbf{a}^{(2)}=f(\mathbf{b}^{(1)})$. Following the input layer, the PIC directly transmits the optically-encoded signal into a hidden layer, composed of another CMXU and six NOFUs, that implements the transformation $\mathbf{a}^{(3)} = f(U^{(2)}\mathbf{a}^{(2)})$. The final layer $U^{(3)}$, implemented with a third CMXU, maps $\mathbf{a}^{(3)}$ to the output $\mathbf{b}^{(3)}$.  Inference  therefore proceeds entirely in the optical domain without  photodiode readout, amplifiers, or digitization between layers. 

An \textit{integrated coherent receiver} (ICR), shown in Figure \ref{architecture}(iv), reads out the amplitude and phase of the DNN output by homodyning each element of the vector $\mathbf{b}^{(3)}$ with a common local oscillator field $E_\mathrm{LO}$. The DNN output is read out by transimpedance amplifiers that convert the  photocurrent vector $\mathbf{i}^{PD}$ to a voltage vector $\mathbf{V}^{PD}$.  $\mathbf{V}^{PD}$ is digitized and then normalized by the sum of voltages measured across all channels $\sum \mathbf{V}^{PD}$ to yield a quasi-probability distribution $\mathbf{V}^{\mathrm{norm}}$ for a classification task. Each sample $\mathbf{x}^{(i)}$ is assigned the label corresponding to the highest probability, i.e.\ $\mathrm{argmax}(\mathbf{V}^{\mathrm{norm}})$.

\begin{figure*}
    \includegraphics[width=7in]{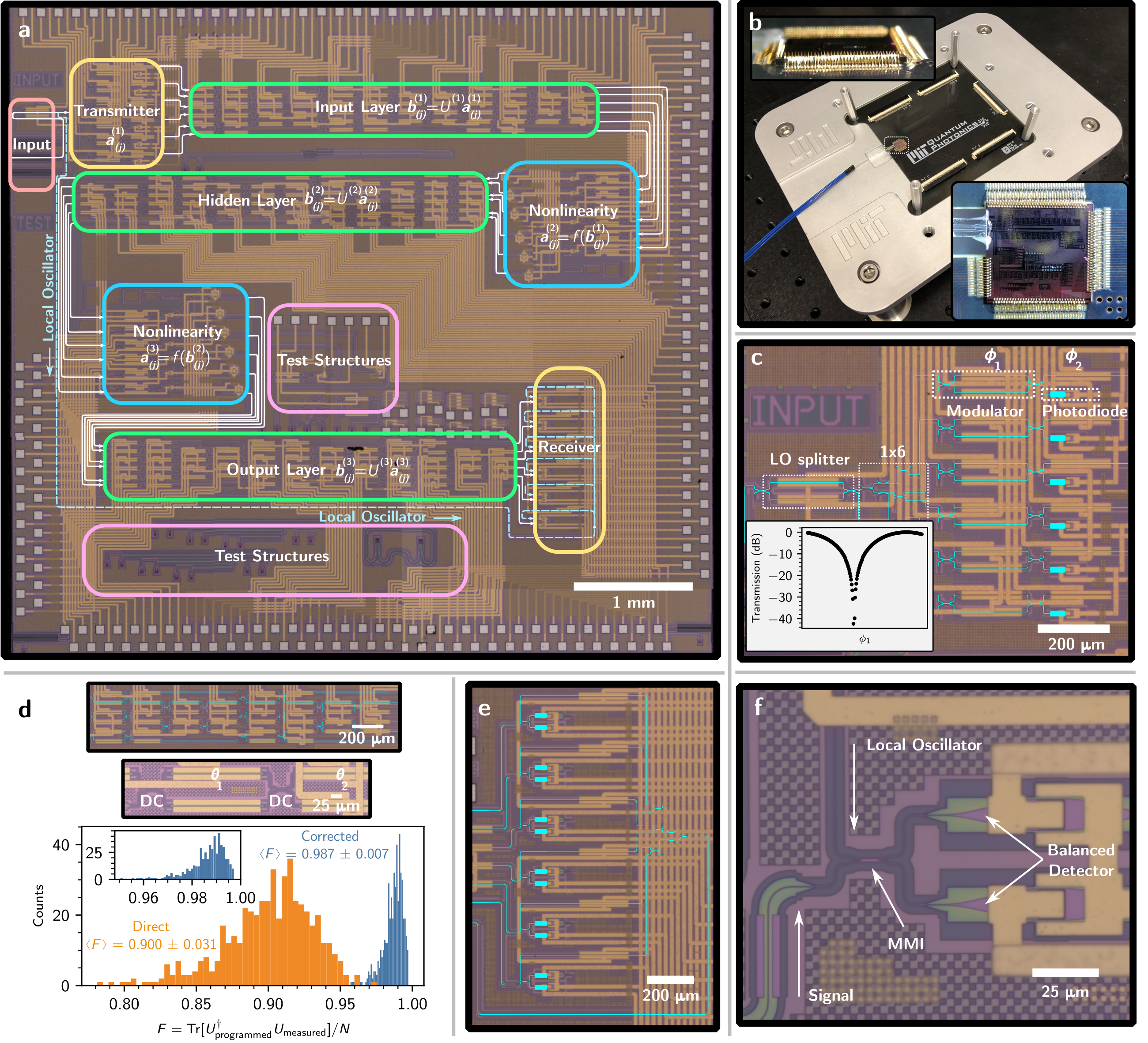}
    \caption{\textbf{a)} Microscope image of the fabricated PIC. Key subsystems of the circuit are highlighted in the same color as the architecture depicted in Figure \ref{architecture}. The signal path through the PIC is indicated in white, while the local oscillator path is outlined in blue. 
    \textbf{b)} Photonic packaging of the PIC for lab testing. Insets show side and top-down views of the packaged PIC.
    \textbf{c)} The fabricated transmitter splits off light coupled into the PIC to a local oscillator and fans out the remainder to six input channels. The inset shows the measured optical response of a typical channel.
    \textbf{d)} The coherent matrix multiplication unit is implemented with a Mach-Zehnder interferometer mesh. Each MZI comprises two directional couplers (DCs), an internal phase shifter $\theta_1$ between the two splitters, and an external phase shifter $\theta_2$ on one output mode. The histogram shows the measured fidelity of 500 arbitrary unitary matrices implemented on a single  layer using a ``direct'' approach (orange) and an approach that takes into account hardware errors and thermal crosstalk (blue).
    \textbf{e)} The integrated coherent receiver (ICR). 
    \textbf{f)} One channel of the ICR. Signal and LO are interfered on a 50-50 MMI and measured using balanced detectors.
    }
    \label{PIC}
\end{figure*}

\section*{Experiment} 

We implemented the FICONN architecture in a commercial silicon photonic foundry process incorporating low-loss edge couplers and waveguides, compact phase shifters, high-speed waveguide-integrated germanium photodiodes, and efficient carrier-based microring modulators. 

Figure 2a shows the PIC, fabricated in a silicon-on-insulator (SOI) process, which monolithically integrates all FICONN subsystems. Demonstrating our system, which required control of 169 active devices and stable optical coupling, necessitated developing a custom photonic package for lab testing. This package, shown in Figure 2b, interfaces the active devices on chip to driver electronics through 236 wirebonds to a printed circuit board. Input light is coupled into the circuit through a single channel of a polarization-maintaining fiber array glued to the chip facet with index-matching epoxy. No light is coupled out of the PIC, as all readout is done on chip with the ICR. 

We measured an end-to-end loss for our system of 10 dB, including 2.5 dB fiber-to-chip coupling loss. As the depth of our system is 91 layers of optical components from input to readout, the end-to-end loss implies a  per-component insertion loss of less than 0.1 dB, enabling single-shot inference across all DNN layers without optical re-amplification.

The key subsystems of the PIC are depicted in Figures 2c$-$f. In Figure 2c, we  show the transmitter for encoding input vectors into the FICONN. 
The light coupled into the chip is first split with an MZI into a local oscillator (LO) path, which is directed to the ICR, and a signal path, which is fanned out to six channels.
Each channel of the transmitter comprises an MZI, which programs the amplitude of one element of $\mathbf{a}^{(1)}_{(j)}$, and a phase shifter on the output that encodes the phase. 
As the inset shows, a typical channel realizes more than 40 dB of extinction, enabling programming of input vectors with more than 13 bits of precision.

Figure \ref{PIC}d shows the coherent matrix multiplication unit (CMXU), which computes linear transformations in the DNN. The CMXU is comprised of a MZI mesh \cite{bogaerts_programmable_2020} of 15 devices, connected in the Clements configuration \cite{clements_optimal_2016}, which implements an arbitrary $6 \times 6$ unitary operation $U^{(1)}$ on the optical fields $\mathbf{a}^{(1)}$. 
As $\mathrm{det}(U^{(1)})=1$, the CMXU conserves optical power in the system with the exception of component insertion losses. Unitary weighting, which redistributes light between optical modes but does not attenuate it, minimizes optical losses and enables single-shot DNN inference without re-amplification or readout between layers. Training unitary layers also avoids the vanishing gradient problem, improving optimization of deep and recurrent neural networks \cite{pmlr-v70-jing17a}. 

We benchmarked the matrix accuracy of the CMXU by programming 500 random $6 \times 6$ unitary matrices sampled from the Haar measure into the device and measuring the fidelity $F = \mathrm{Tr}[U^\dagger_\mathrm{programmed} U_\mathrm{measured}]/6$. 
In the histogram in Figure \ref{PIC}d, we show the measured fidelity obtained with a ``direct'' programming, where we algorithmically decompose the phase shifter settings as outlined in \cite{clements_optimal_2016}, and using a modified programming that corrects for hardware errors, losses, and thermal crosstalk \cite{Bandyopadhyay:21, hamerly_stability_2021, hamerly_accurate_2021}. While a direct programming only achieves a matrix fidelity of $\langle F \rangle = 0.900 \pm 0.031$, correcting for hardware non-idealities improves this value to $\langle F \rangle = 0.987 \pm 0.007$ for the CMXU. To the best of our knowledge, this is the highest reported fidelity for a programmable photonic matrix processor.

Figure \ref{PIC}e shows the integrated coherent receiver (ICR), which measures the amplitude and phase of the output signal $\mathbf{b}^{(3)}$ of the DNN. Each  channel, as shown in Figure \ref{PIC}f, interferes the  signal field with the LO using a 50-50 multimode interferometer (MMI) and measures the outputs with a pair of balanced detectors. A phase shifter is used to select the quadrature being read out.

\begin{figure*}
    \includegraphics[width=7in]{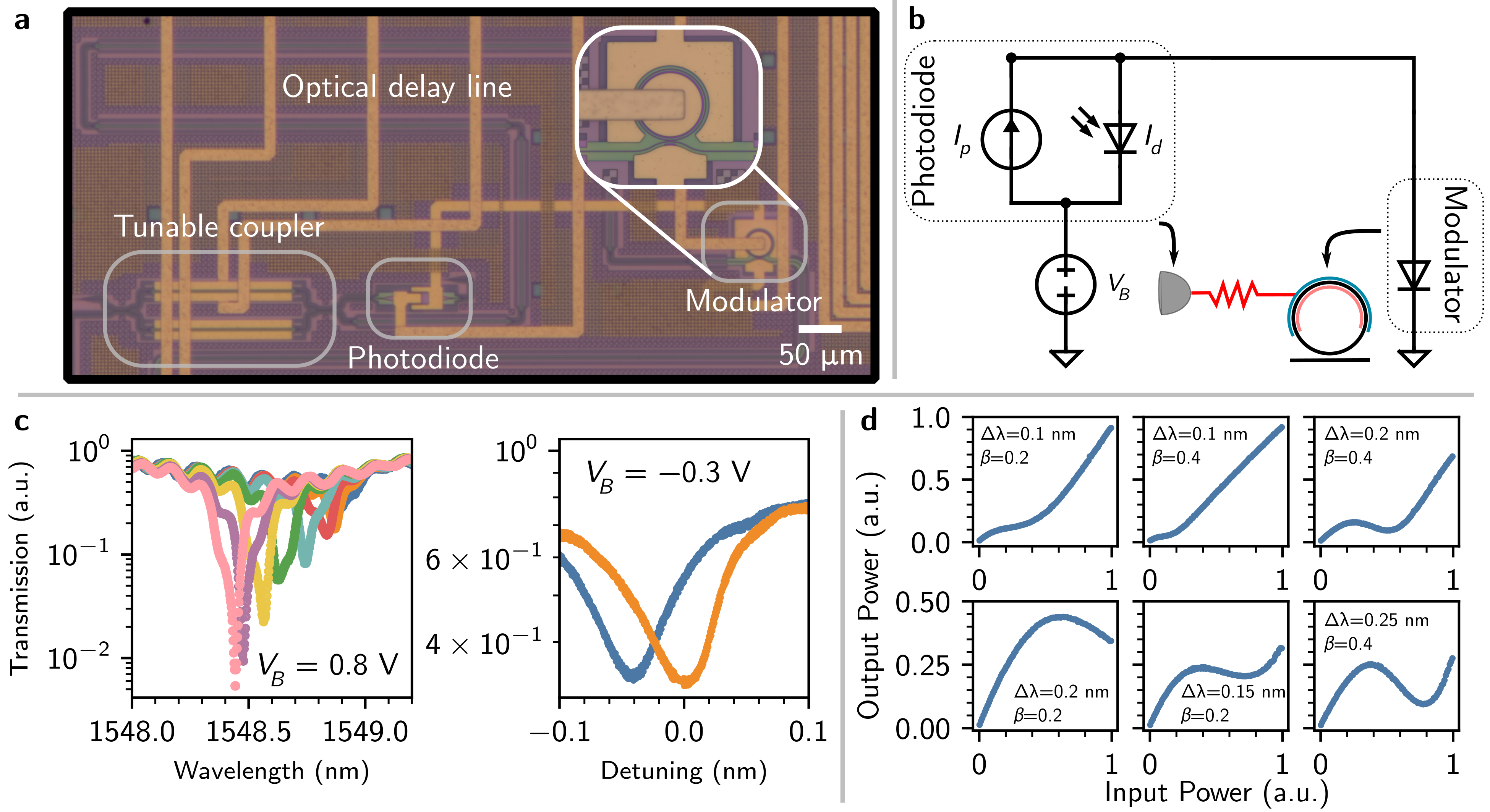}
    \caption{\textbf{a)} The fabricated NOFU. A programmable MZI determines the fraction of light tapped off to the photodiode, and a waveguide delay line synchronizes the optical and electrical pulses. A \textit{pn}-doped microring resonator modulates the incident field. \textbf{b)} Circuit diagram of resonant EO nonlinearity. The photocurrent $I_p$ directly drives a \textit{pn}-doped resonant modulator. No amplifier stage is required between the two and the devices are directly connected on chip. By adjusting the bias voltage $V_B$, the nonlinearity can be operated in forward or reverse bias. \textbf{c)} Left: Detuning of the cavity resonance at various incident optical powers when operated in carrier injection mode ($V_B > 0$). Right: Cavity detuning in carrier depletion mode ($V_B < 0$). Our system realizes close to a linewidth detuning without the use of any amplifier, improving energy consumption and latency of the nonlinearity. A full linewidth detuning can be realized by further engineering the cavity finesse. \textbf{d)} Activation functions measured on chip. Arbitrary function shapes can be realized by adjusting the cavity detuning $\Delta \lambda$ and fraction of light $\beta$ tapped off to the photodiode.}
    \label{activation}
\end{figure*}

The programmable nonlinear optical function unit (NOFU) is shown in Figure 3. To realize a programmable coherent optical activation function, we developed the resonant electro-optical nonlinearity shown schematically in Figure \ref{architecture}iii).
This device directs a fraction $\beta$ of the incident optical power $|b|^2$ into a photodiode by programming the phase shift $\theta$ in an MZI. The photodiode is electrically connected to a $pn$-doped resonant microring modulator, and the resultant photocurrent (or photovoltage) detunes the resonance by either injecting (or depleting) carriers from the waveguide. The remainder of the incident signal field passes into the microring resonator; the nonlinear modulation of the electric field $b$ by the cavity, which is dependent on the incident optical power $|b|^2$, results in a coherent nonlinear optical function for DNNs. Setting the detuning of the cavity and the fraction of optical power tapped off to the photodiode determines the  implemented function.

Figure \ref{activation}a shows the fabricated device, where the photodiode output is directly connected on chip to the modulator. An integrated heater aligns the microring resonance to the programmed detuning, and an optical delay line placed between the tunable coupler and modulator synchronizes the optical and electrical pulses. 

\begin{figure*}
    \includegraphics[width=7in]{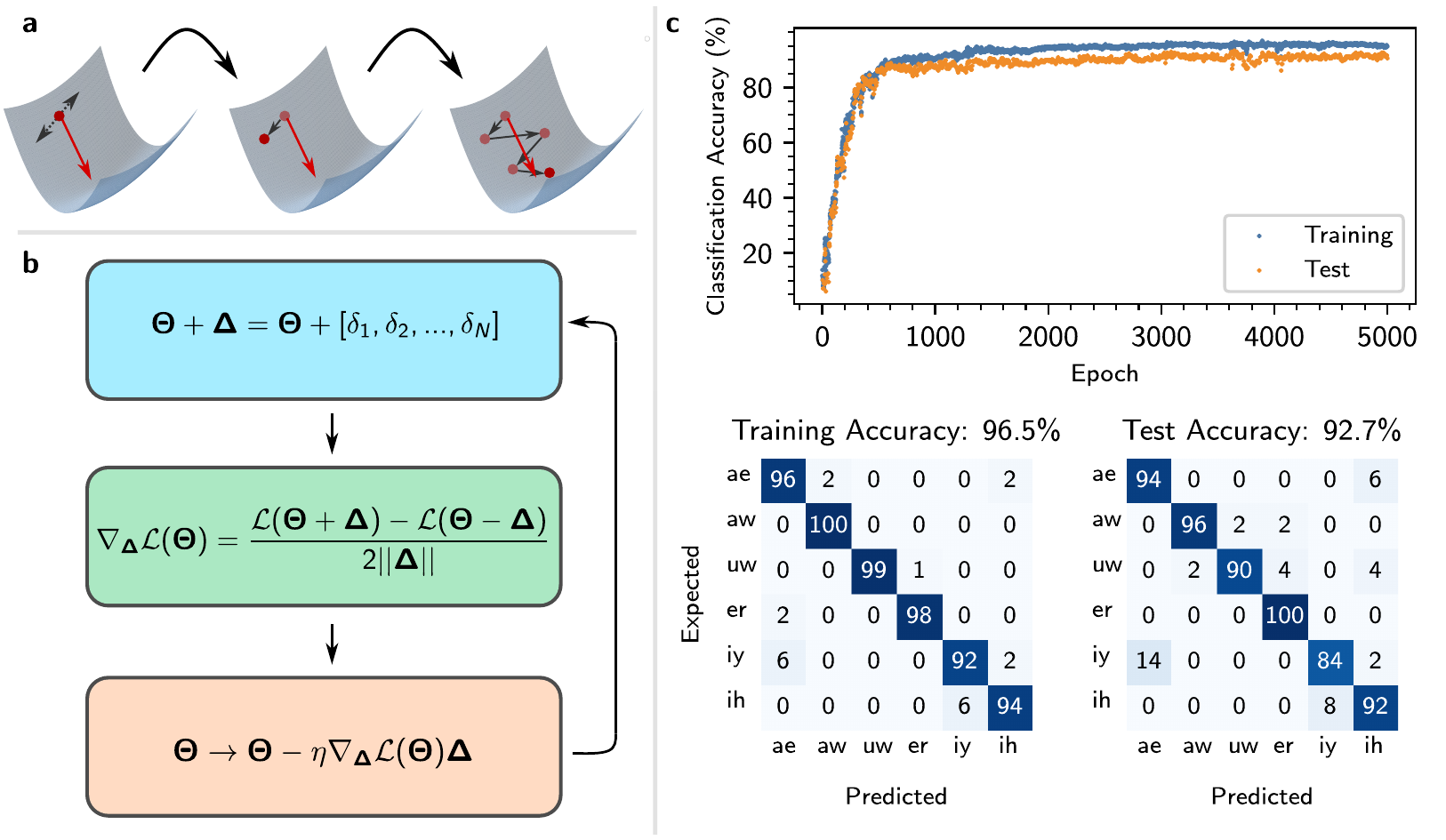}
    \caption{\textbf{a)} A multivariate cost function $\mathcal{L}(\mathbf{\Theta})$ can be minimized by computing the directional derivative of the function along a random direction (black). This directs the optimization along the component of the gradient (red) parallel to the search direction. Over multiple iterations, the steps taken along random directions average to follow the direction of steepest descent to the minimum.  \textbf{b)} \textit{In situ} training procedure. At every iteration, the directional derivative of the cost function $\mathcal{L(\mathbf{\Theta})}$ is computed in hardware along a randomly chosen direction $\mathbf{\Delta}$ in the search space. $\mathbf{\Delta}$ is chosen from a Bernoulli distribution to be $\pm \delta$. The weights $\mathbf{\Theta}$ are then updated by the measured derivative following a learning rate $\eta$ chosen as a hyperparameter of the optimization. \textbf{c)} \textit{In situ} training of a photonic DNN for vowel classification. We obtain 92.7\% accuracy on a test set, which is the same as the performance (92.7\%) obtained on a digital model with the same number of weights. Despite not having direct access to gradients, our approach produces a training curve similar to those produced by standard gradient descent algorithms.}
    \label{training}
\end{figure*}

The electrical circuit for the NOFU is shown in Figure \ref{activation}b. Incident light generates a reverse current in the photodiode; depending on the bias voltage $V_B$, this either injects carriers into the modulator or generates a photovoltage that depletes the modulator of carriers. 
Figure \ref{activation}c shows the device response in injection (left) and depletion modes (right). In injection mode, optical power modulates both the loss and phase of the resonator, producing a strong nonlinear response to the incident field $b$. 
In depletion mode, we observe nearly a linewidth detuning when the incident light is switched on vs.\ off, which is induced by the voltage produced by the photodiode. 

The NOFU is designed to implement programmable nonlinear activation functions at high speeds with ultra-low energy consumption. 
This required separately optimizing the cavity parameters, which determine the microring response time, and closely integrating the photodiode and modulator together on the PIC to minimize total device capacitance, and therefore the $RC$ time delay. 
In injection mode, we found that 75~$\mu$A photocurrent was sufficient to detune the resonator by a linewidth. 
As each NOFU performs the equivalent of two multiplications in digital electronics, over a carrier lifetime of $\sim$1 ns this corresponds to an energy consumption of 30 fJ per nonlinear operation (NLOP). 

Compared to prior approaches \cite{williamson_reprogrammable_2020, ashtiani_-chip_2022}, the NOFU directly drives the modulator through the photodiode and eliminates the amplifier stage between them. This greatly improves the latency and energy efficiency of the device, as high speed transimpedance amplifiers consume hundreds of milliwatts of power \cite{ahmed_34gbaud_2018}. For our device, incorporating such an amplifier would have increased the power consumption by two orders of magnitude to about 3 pJ/NLOP. Our design, which eliminates intermediate amplifier circuitry and is therefore ``receiverless'' \cite{miller_attojoule_2017}, is not only more energy-efficient, but also eliminates the latency introduced by the amplifier.

In Figure \ref{activation}d, we show several of the activation functions measured on chip. The programmability of the device enables a wide range of nonlinear optical functions to be realized. By tuning the fraction of power tapped off to the photodiode and the relative detuning of the cavity, we can not only program the form of the nonlinear function, but also train it during model optimization.

\section*{\textit{In situ} training}

The accuracy of the inference output depends on the model parameters $\mathbf{\Theta}$ of the FICONN system, comprising a total of $N_\mathrm{model}= N_\mathrm{layer} N_\mathrm{neuron}^2 + 2N_\mathrm{neuron}(N_\mathrm{layer}-1)   = 132$ real-valued phase shifter settings controlled with 16 bits of precision. These parameters can either be determined offline by training on a digital computer \cite{shen_deep_2017, sludds_delocalized_2022, bernstein_single-shot_2022}, using a digital model of the hardware \cite{wright_deep_2022}, or by training the hardware parameters \textit{in situ}. 

Training \textit{in situ} takes advantage of low latency inference on optical hardware, reducing the time and energy required for model optimization. Previous work on \textit{in situ} training has focused on developing optical implementations of ``backpropagation,'' which is the standard for training electronic DNNs \cite{hughes_training_2018, pai2022}. However, these approaches train only the linear layers of a photonic system and require evaluating gradients of activation functions on a digital system, thereby  limiting the optical acceleration obtained by computing a multi-layer DNN in a single shot.
Alternatively, genetic algorithms have been used to optimize weights on chip \cite{zhang_efficient_2021}, but they are challenging to scale to large model sizes and require many generations to converge.

We trained the model parameters of the FICONN \textit{in situ}, including those of the activation functions, by evaluating the derivatives of those parameters directly on the hardware. To the best of our knowledge, this is the first demonstration of \textit{in situ} training of a photonic DNN. Our approach, which is based on prior work on \textit{in situ} optimization of analog VLSI neural networks \cite{Cauwenberghs_1992, spall_1998}, is robust to noise, performs gradient descent on average, and is guaranteed to converge to a local minimum.
Moreover, it is not limited to our specific system, but can be generalized to any hardware architecture for photonic DNNs.

A direct approach to computing the gradient on hardware would be to perturb the model parameters $\mathbf{\Theta} = [\Theta_1, \Theta_2, ..., \Theta_N]$ one weight at a time and repeatedly batch the training set through the system \cite{shen_deep_2017}. This procedure produces a forward difference estimate of the loss gradient $\nabla \mathcal L(\mathbf{\Theta})$ with respect to all weights. Moreover, since the derivatives are evaluated directly on chip, this procedure extends to other hardware parameters, such as the detuning and fraction of power tapped off in the NOFU. The drawback to this approach is that for $N$ parameters, it requires batching the training set through the hardware $2N$ times. 

Our approach  varies all model parameters $\mathbf{\Theta}$ simultaneously. Figure \ref{training}b sketches the optimization procedure. Instead of perturbing the parameters one weight at a time, during training the system perturbs all parameters towards a random direction $\mathbf{\Delta}$ in search space, i.e. $\mathbf{\Theta} \rightarrow \mathbf{\Theta} + \mathbf{\Delta}=\mathbf{\Theta} + [\delta_1, \delta_2, ..., \delta_N]$. At each iteration the system then computes the directional derivative:
\begin{equation}
    \nabla_\mathbf{\Delta} \mathcal{L}(\mathbf{\Theta}) =
    \frac{\mathcal{L}(\mathbf{\Theta} + \mathbf{\Delta})-\mathcal{L}(\mathbf{\Theta} - \mathbf{\Delta})}{2||\mathbf{\Delta}||}
\end{equation}
As in standard gradient descent, the weights $\mathbf{\Theta}$ are then updated to $\mathbf{\Theta} \rightarrow \mathbf{\Theta} - \eta \nabla_\mathbf{\Delta} \mathcal{L}(\mathbf{\Theta}) \mathbf{\Delta}$, where $\eta$ is a learning rate chosen as a hyperparameter of the system. 

Compared to the forward difference approach outlined earlier, our approach  requires batching the training set through the hardware only twice per iteration. Moreover, we obtain true estimates of the cost function $\mathcal L$ and the derivative $\nabla_\mathbf{\Delta} \mathcal{L}(\mathbf{W})$, ensuring that component errors or errors in calibration do not affect the accuracy of training.
Unlike other derivative-free optimization methods, our approach will always track the direction of steepest descent, as errors in the gradient direction average out to zero over multiple epochs \cite{Cauwenberghs_1992, spall_1998} (see Supplementary Information [SI]). 

We implemented \textit{in situ} training of $\mathbf{\Theta}$, which includes weights and nonlinear function parameters, for a standard vowel classification task (dataset available at \cite{vowels}). 
At each epoch, we batched a training set of 540 samples into the system and implemented the optimization loop described in Figure \ref{training}b with a learning rate $\eta = 0.002$. We reserved part of the data ($N = 294$) to evaluate the trained model on inputs it had not seen before.

The top plot of Figure \ref{training}c shows the classification accuracy of both datasets during training. Our system achieves over 96\% accuracy on the training set, and over 92\% accuracy on the test set, as shown in the confusion matrices at the bottom. When training a digital system, we found it also obtained the same accuracy on the test set. Each epoch batches the training set only three times through the system; two times to evaluate the derivative $\nabla_\mathbf{\Delta} \mathcal{L}(\mathbf{\Theta})$ and once more to evaluate $\mathcal L(\mathbf{\Theta})$ at the current parameter set $\mathbf{\Theta}$. We observed that the system quickly trained to an accuracy exceeding 80\%, and then slowly asymptoted to a training accuracy of 96\%. This behavior resembles the optimization trajectories of other first-order methods for training DNNs in electronics, such as stochastic gradient descent. Moreover, our system successfully trains using only 16-bit accuracies for the weights. Lower precision weights reduce memory requirements for training; however, digital systems are challenging to train with fewer than 32 bits due to numerical errors in gradients accumulating during backpropagation \cite{micikevicius2018mixed}. 

\section*{Discussion}

An important DNN metric is the latency $\tau_\text{latency}$ of inference, i.e. the time delay between input of a vector and the DNN output. For the FICONN, $\tau_\text{latency}$ is dominated by the optical propagation delay, which we estimate from the PIC subsystems (as described in the SI) as $3\tau_\text{CMXU} + 2\tau_\text{NOFU} + \tau_\text{TX to U1} + \tau_\text{U3 to RX} + \tau_\text{U-turn} \approx$ 435 ps.

Each inference requires $N_\text{OPS}\approx 2 M N^2$ matrix operations (see SI), where $M$ is the number of layers and $N$ the number of modes.  Dividing the FICONN's energy consumed during $\tau_\text{latency}$ by $N_\text{OPS}$ upper-bounds the energy-per-operation as   
\begin{equation}
E_\text{OP} \approx \frac{\tau_\text{latency}}{2} \left [P_\text{PS} + \frac{P_\text{NOFU}}{N} + \frac{P_\text{TX} + P_\text{ICR}}{MN} \right ],
\label{energy}
\end{equation} 
where $P_j$ denotes the power dissipation of subsystem $j$. In the SI, we estimate upper bounds to the on-chip energy consumption of 9.8 pJ/OP and a throughput of $N_\text{OPS}/\tau_\text{latency}\approx 0.53$ tera-operations per second (TOPS) per inference.  

\begin{table}[h]
\begin{tabular}{|c|c|c|c|c|c|}
$N$                 & Phase shifter             & $E_\text{OP}$ & $E_\text{total,est}$ & $\tau_\text{latency}$ & TOPS  \\
\hline
6 & \vtop{\hbox{\strut Thermal}\hbox{\strut (this work)}} & 9.8 pJ/OP & 11.7 pJ/OP         & 435 ps  & 0.53  \\
6 & \vtop{\hbox{\strut Undercut}\hbox{\strut thermal \cite{dong_thermally_2010}}}  & 35 fJ/OP     & 546 fJ/OP   & 140 ps  & 12    \\
6 & MEMS \cite{baghdadi_dual_2021, gyger_reconfigurable_2021}               & 1.6 fJ/OP  &  513 fJ/OP     & 140 ps  & 12    \\
64 & MEMS \cite{baghdadi_dual_2021, gyger_reconfigurable_2021}               & 0.84 fJ/OP    &   54 fJ/OP   & 1.4 ns  & 1240  \\
128 & MEMS \cite{baghdadi_dual_2021, gyger_reconfigurable_2021}              & 0.79 fJ/OP  &  27 fJ/OP      & 2.7 ns  & 4940  \\
256 & MEMS  \cite{baghdadi_dual_2021, gyger_reconfigurable_2021}             & 0.77 fJ/OP   &  14 fJ/OP     & 5.4 ns  & 19700
\end{tabular}
\caption{Performance metrics for a three-layer FICONN with $N$ neurons. We list the on-chip energy consumption $E_\text{OP}$, as well as an estimate of the total power dissipation $E_\text{total,est}$ including optimized driver electronics. The predicted metrics assume inference on large batches of vectors with  resonant modulators at 50 GHz \cite{timurdogan_ultralow_2014}. For latency, we assume a device length of 500~$\mu$m and an optimized layout, while our reported latency uses the actual waveguide layout fabricated on the PIC.}
\label{table}
\end{table}

The FICONN's power consumption is dominated by the thermal phase shifters, which require $\sim$25 mW of electrical power to produce a $\pi$ phase shift. Table \ref{table} lists the key parameters of our proof-of-concept FICONN (top row), along with estimates for alternative published phase shifter technologies for varying $N$ and $M=3$. These estimates suggest that low-power quasi-static phase shifters in combination with high-speed modulators \cite{timurdogan_ultralow_2014} could push total energy consumption to $\sim 10$ fJ/OP for large systems, while maintaining ns latencies and throughputs of thousands of TOPS. In comparison, systolic arrays such as the tensor processing unit (TPU) require at minimum $N+1$ clock cycles for a single $N \times N$ matrix-vector multiplication. A three-layer DNN with $N=256$ neurons would require $\sim$1 $\mu$s to compute at a 700 MHz clock speed \cite{Jouppi2017}, which is more than two orders of magnitude longer than in a photonic processor.

The ultra-low latency of inference in the FICONN could greatly improve the speed of training models, which consumes significant energy \cite{Strubell_Ganesh_McCallum_2020} and has motivated a search for efficient scheduling algorithms that reduce  training time \cite{You_ImageNet_2018}. \textit{In situ} training could also improve the performance of DNN models, as training with noise has been suggested to regularize models,  preventing overfitting \cite{Camuto_Explicit_2020} and improving their adversarial robustness \cite{Liu_2018_ECCV} to small changes in the input. This regularization can be implemented automatically by leveraging quantum noise in hardware. We observed this effect in our own experiments; while both the FICONN and a digital system obtained similar performance on the test set for the classification task studied, the digital system overfit the model, achieving perfect accuracy on the training set (see SI). Finally, our implementation of \textit{in situ} training, which does not require a digital system for computing gradients, is compatible with feedback-based ``self-learning'' photonic devices \cite{feldmann_all-optical_2019, Marquadt_2021}, enabling fast, autonomous training of models without any required external input.

The FICONN, which is implemented in a foundry-fabricated photonic integrated circuit, could be scaled to larger sizes with current-day technologies. Silicon photonic foundries have already produced functional systems of up to tens of thousands of components \cite{sun_large-scale_2013}. Spectral multiplexing, for instance through integration of microcomb sources with silicon photonics \cite{shu_microcomb-driven_2022}, could enable classifying data simultaneously across many wavelength channels, further reducing energy consumption and increasing throughput.  The system's energy consumption would further improve by optimization of the NOFU; while our implementation makes use of microring resonators, photonic crystal modulators \cite{nozaki_femtofarad_2019}, microdisks \cite{timurdogan_ultralow_2014}, or hybrid integration of lithium niobate \cite{li_all-optical_2022, wang_integrated_2018} would further reduce the activation function to less than 1 fJ/NLOP. 

Our implementation of the FICONN makes use of feedforward unitary circuits, which implement fully-connected layers in a DNN. However, this architecture can be generalized to other types of neural networks. For example, temporal or frequency data may be classified using recirculating waveguide meshes \cite{perez-lopez_multipurpose_2020}, which can implement feedback and resonant filters. Such a system, where phase shifter settings are trained \textit{in situ} \cite{perez-lopez_multipurpose_2020, mak_wavelength_2020}, may be used for intelligent processing of microwave signals in the optical domain.

\section*{Conclusion}

We have demonstrated a  coherent optical DNN on a single chip that performs both inference and \textit{in situ} training. The FICONN system introduces inline nonlinear activation functions based on modulators driven by ``receiverless'' photodection, eliminating the latency and power consumption introduced by optical-to-electrical conversion between DNN layers and preserving phase information for optical data to be processed coherently. The system fabrication relied entirely on commercial foundry photolithography, potentially enabling scaling to wafer-level systems. Scaling these systems up to hundreds of modes would lower energy consumption to $\sim$10 fJ/OP, while maintaining latencies  orders of magnitude lower than electronics.

Moreover, we have demonstrated \textit{in situ} training of DNNs by estimating derivatives of model parameters directly on  hardware. Our approach is also generalizable to other photonic DNN hardware being currently studied. \textit{In situ} training, which takes advantage of the optically-accelerated forward pass enabled by receiverless hardware, opens the path to a new generation of devices that learn in real time for sensing, autonomous driving, and telecommunications.

\section*{Methods}
{\footnotesize\par}

\noindent\footnotesize{}\textbf{Photonic integrated circuit.} The photonic integrated circuit (PIC) was fabricated in a commercial silicon photonics process by Elenion Technologies. Waveguides were defined in a  crystalline silicon layer cladded by silicon dioxide, and the optical signals were routed with partially-etched waveguides to minimize propagation loss and backscattering. Most signal routing was done with single-mode waveguides, while longer distance propagation used multimode waveguides to further reduce transmission losses. Input light ($\lambda = 1564$ nm) was edge coupled into the chip, while output signals were measured on chip with waveguide-integrated germanium photodiodes. Mach-Zehnder interferometers were programmed using 200 $\mu$m long thermal phase shifters, which induce a refractive index change by locally heating the waveguide. The nonlinear optical function unit was realized with a 20 $\mu$m radius microring resonator where the waveguide core is \textit{pn} doped.

Light was coupled into the system through a polarization-maintaining fiber array glued to the chip facet. The PIC was bonded to a copper plane on a printed circuit board for heatsinking and electrically driven through 236 wirebonds.  We thermally stabilized the system using a Peltier module connected to a precision feedback system that locked the chip temperature to within 31$^\circ$ $\pm$ 0.004$^\circ$ C (Arroyo Instruments 5400).

Devices on the PIC were electrically controlled through a 192-channel software programmable current source (Qontrol Systems Q8iv). Each channel sources up to 24 mA of current with 16 bits of precision, corresponding to approximately 0.4 mrad precision in our system. For faster transmission of input vectors into the DNN, we designed a custom 16-bit current driver system that used a microcontroller to buffer the training set in memory, which enabled training at the maximum DAC speed rather than the speed of the serial connection to the computer. This system was paired with a custom receiver board, which read out the  photodiodes on chip with a transimpedance amplifier and 18-bit ADC.\\

\noindent\footnotesize{}\textbf{System characterization.} Light coupled into the PIC was split into local oscillator (LO) and signal paths with a programmable MZI. The LO is directed to the ICR, while the signal is fanned out to the six channels of the transmitter through a splitting tree of 50-50 multimode interferometers (MMIs). Each channel of the transmitter was calibrated using a photodiode on the drop port of the MZI. For each mode, we swept the current $I$ driven into the thermal phase shifter and measured the output transmission $T(I)$. To produce a mapping between current and phase for an MZI, we fit the expression $A \pm B \cos (p_4 I^4 + p_3 I^3 + p_2 I^2 + p_1 I + p_0)$ to $T(I)$. The sign of the expression depends on whether transmission is measured at the cross ($+$) or bar ($-$) port.

The first two meshes were calibrated at their output using the photodiodes that drive the NOFUs, while the final mesh was calibrated with the receiver. To characterize an uncalibrated mesh, we began by transmitting light down the main diagonal. Light was transmitted into input 1 and the internal phase shifters along the main diagonal were optimized in a round robin fashion to maximize the signal at output 6. This procedure deterministically initialized the main diagonal to the ``cross'' ($\theta_1 = 0$) state, as there is only one possible path in the circuit between the two modes. The devices were then characterized by routing light down diagonals of the circuit. External phase shifters were calibrated by programming ``meta-MZIs'' into the device, as described in \cite{prabhu_accelerating_2020}. 

To correct for hardware errors, we programmed 300 Haar random unitaries into the mesh and measured the output from transmitting 100 random input vectors. The measured data was fit to a software model of an MZI mesh that incorporates the effects of beamsplitter errors, waveguide losses, and thermal crosstalk. The software model's parameters were optimized to fit the measured data using the limited-memory Broyden–Fletcher–Goldfarb–Shanno (L-BFGS) algorithm. 

We found the software model is able to predict the hardware outputs with an average fidelity of 0.969 $\pm$ 0.023. Hardware error correction was implemented by determining the required hardware settings to implement a desired matrix within the software model and then porting them to the chip. Similar to our previous work \cite{Bandyopadhyay:21}, our approach here efficiently corrects for component errors, as no real-time optimization is done on the hardware. However, fitting the device response to a software model eliminates the need to calibrate component errors one at a time.

The fidelity results shown in Fig.\ \ref{PIC} were obtained by programming random unitary matrices $U_\text{programmed}$ sampled from the Haar measure into the circuit and sequentially transmitting the columns of $U^\dagger_\text{programmed}$ to compute the metric $F = \mathrm{Tr}[U^\dagger_\text{programmed} U_\text{hardware}]/N$. As the inverse of a unitary matrix is its adjoint, for a perfect hardware implementation of $U_\mathrm{programmed}$ the quantity $F$ should equal 1. 

Insertion losses in the CMXU were inferred by transmitting light down different paths in the circuit and fitting the measured photocurrent to the number of devices light passed through. We measured the insertion loss per MZI to be $0.22 \pm 0.05$ dB, corresponding to a loss per CMXU of $1.32 \pm 0.30$ dB. 

The wavelength spectra shown in Figure \ref{activation}c were measured on a test structure of the NOFU by varying the incident optical power. We measured the microring to have a quality factor $Q = 8300$ when no current is injected into the device, which corresponds to a cavity lifetime $\tau_\text{NL} = Q/\omega$ of 6.6 ps. The activation functions measured in Figure \ref{activation}d were obtained with the integrated coherent receiver on the PIC. To operate in injection mode, $V_B$ must be sufficiently high to ensure the modulator can be forward biased ($\sim$0.7 V). Otherwise, the device operates in photovoltaic mode and generates a reverse bias $\sim$0.3 V across the modulator, which increases the width of the junction depletion zone and removes carriers from the waveguide.
\\

\noindent\footnotesize{}\textbf{\textit{In situ} training.} For demonstrating \textit{in situ} training, we used vowel classification data from the Hillenbrand database (available at \cite{vowels}). We used the first three formants $F_1, F_2, F_3$ at steady state and at 50\% of the vowel duration as the six input features for each datapoint. Each input vector was normalized to ensure that the maximum value was 1. We used 540 samples for training and evaluated the performance of the model on a test set of 294 samples.

We initialized the weights for the unitary layers randomly over the Haar measure \cite{pai_matrix_2019}. At every epoch, we performed the following optimization loop:
\begin{enumerate}
    \item Perturb the system parameters $\mathbf{\Theta}$ by a random displacement $\pm \mathbf{\Delta}$. $\mathbf{\Delta}$ is a vector of the same length as $\mathbf{\Theta}$ whose elements are chosen from a Bernoulli distribution to be $\pm \delta$. $\delta$ is a hyperparameter of the optimization; we used $\delta = 0.05$ in our experiments.
    \item Batch the training set through the system and compute the loss $\mathcal L = \sum_j \mathbf{y}_{(j)}^\text{train} \log \mathbf{V}_{(j)}^\text{norm}$ for hardware parameters $\mathbf{\Theta} \pm \mathbf{\Delta}$.
    \item Estimate the directional derivative $\nabla_\mathbf{\Delta} \mathcal{L}(\mathbf{\Theta})$ along $\mathbf{\Delta}$ and update the hardware parameters $\mathbf{\Theta} \rightarrow \mathbf{\Theta} - \eta \nabla_\mathbf{\Delta} \mathcal{L}(\mathbf{\Theta})\mathbf{\Delta}$. We found that $\eta = 0.002$ provided stable convergence.
\end{enumerate}
During training a software feedback loop stabilized the power coupled into the chip, as variations in optical power affected the response of the electro-optic nonlinearity. 

The \textit{in situ} training experiments were conducted with the NOFU in injection mode ($V_B = 0.8$ V) due to the wider range of nonlinear functions we could realize on chip. We optimized our device design for carrier injection, which accounts for the comparatively lower efficiency in depletion mode. However, even our non-optimized design realizes nearly a full linewidth detuning in depletion; therefore, we expect a modest improvement in the cavity finesse would be sufficient to realize full modulation in future iterations.
\\

\noindent\footnotesize{}\textbf{Data availability.} The data that support the plots in this paper are available from the corresponding authors upon reasonable request.\\

\noindent\footnotesize{}\textbf{Code availability.} The code used to generate the results of this paper is available from the corresponding authors upon reasonable request.\\

\noindent\footnotesize{}\textbf{Acknowledgments.} S.B. was supported by a National Science Foundation (NSF) Graduate Research Fellowship under grant no. 1745302, NSF awards 1839159 (RAISE-TAQS) and 2040695 (Convergence Accelerator), and the Air Force Office of Scientific Research (AFOSR) under award number FA9550-20-1-0113. A.S. was also supported by an NSF Graduate Research Fellowship and the aforementioned AFOSR award, as well as NSF award 1946976 (EAGER) and NTT Research.\\
\\
\noindent The authors would like to acknowledge Paul Gaudette and Dr. David Scott of Optelligent for packaging the photonic integrated circuit; Dr.\ Ruizhi Shi and Dr.\ Hang Guan for feedback on the photonics layout; Dr.\ Sri Krishna Vadlamani for discussions on hardware-aware training; Liane Bernstein for discussions on DNN applications and feedback on the manuscript; Dr.\ Jacques Carolan and Mihika Prabhu for discussions on chip packaging and testing of the photonics; and Dave Lewis for assistance with the use of machining tools.\\ 

\noindent\footnotesize{}\textbf{Competing interests.} S.B., R.H., and D.E.\ have filed US Patent Applications 17/556,033 and 17/711,640 on error correction algorithms for programmable photonics. 
N.H.\ is CEO of Lightmatter.
D.B.\ is Chief Scientist at Lightmatter. 
M.S.\ is VP, Packaging, Photonics, \& Mixed-Signal at Luminous Computing.
M.H.\ is President of Luminous Computing. 
D.E.\ holds shares in Lightmatter, but received no support for this work. 
The other authors declare no competing interests.\\

\noindent\footnotesize{}\textbf{Author contributions.} S.B.\ and D.E.\ conceived the experiments. S.B.\ designed the photonic integrated circuit, chip packaging, and control electronics, calibrated the system, and conducted the experiments. A.S.\ assisted with characterizing the electro-optical nonlinearity. S.B., S.K., and D.E.\ developed the \textit{in situ} training scheme. R.H.\ assisted with developing calibration procedures for the system and interpreting the results of the \textit{in situ} training experiments. N.H.\ and D.B.\ architected the photonic integrated circuit. D.B. performed preliminary evaluation of the PIC in Tensorflow. M.S.\ and M.H.\ fabricated the photonic integrated circuit. S.B.\ and D.E.\ wrote the manuscript with input from all authors.\\
\\\noindent\footnotesize{}\textbf{Supplementary information} is available for this paper.\\
\\
\noindent\footnotesize{}\textbf{Correspondence and requests for materials} should be addressed to Saumil Bandyopadhyay or Dirk Englund.

\bibliographystyle{naturemag}
\bibliography{Bibliography}

\end{document}

\setcounter{page}{12}

\title{Supplementary Information: Single chip photonic deep neural network with accelerated training}
\author{Saumil Bandyopadhyay}
\email{saumilb@mit.edu}
\affiliation{Research Laboratory of Electronics, MIT, Cambridge, MA 02139, USA}

\author{Alexander Sludds}
\affiliation{Research Laboratory of Electronics, MIT, Cambridge, MA 02139, USA}

\author{Stefan Krastanov}
\affiliation{Research Laboratory of Electronics, MIT, Cambridge, MA 02139, USA}

\author{Ryan Hamerly}
\affiliation{Research Laboratory of Electronics, MIT, Cambridge, MA 02139, USA}
\affiliation{NTT Research Inc., PHI Laboratories, 940 Stewart Drive, Sunnyvale, CA 94085, USA}

\author{Nicholas Harris}
\affiliation{Research Laboratory of Electronics, MIT, Cambridge, MA 02139, USA}

\author{Darius Bunandar}
\affiliation{Research Laboratory of Electronics, MIT, Cambridge, MA 02139, USA}

\author{Matthew Streshinsky}
\affiliation{Nokia Corporation, New York, NY, 10016, USA}

\author{Michael Hochberg}
\affiliation{Luminous Computing Inc., Mountain View, CA, 94041, USA}

\author{Dirk Englund}
\email{englund@mit.edu}
\affiliation{Research Laboratory of Electronics, MIT, Cambridge, MA 02139, USA}

\maketitle
\onecolumngrid
\tableofcontents

\newpage
\section{Device characterization}
\subsection*{Transmitter}
A Mach-Zehnder interferometer (MZI) performs the programmable $2\times2$ unitary operation:
\begin{equation}
U(\theta_1, \theta_2) = 
ie^{i \theta_1/2}
\begin{bmatrix}
e^{i\theta_2}  \sin(\theta_1/2) & e^{i\theta_2} \cos(\theta_1/2) \\
\cos(\theta_1/2) & -\sin(\theta_1/2) \\
\end{bmatrix}
\end{equation}
To characterize a single MZI, we first input light into one port of the device and measure the output transmission $T(\theta_1) = P_\text{out} / P_\text{in}$. For an ideal device, the output transmission at the bar port is
\begin{equation}
    T_{\rm bar}(\theta_1) = \sin^2 \left ( \frac{\theta_1}{2} \right)
\end{equation}
and at the cross port:
\begin{equation}
    T_{\rm cross}(\theta_1) = \cos^2 \left ( \frac{\theta_1}{2} \right)
\end{equation}

In a fabricated MZI, $\theta_1$ is determined by the total dissipated power $I \times V(I)$, where $I$ is the programmed current and $V(I)$ is the voltage dropped across the device. To characterize a device in the transmitter, where the cross port of each channel has a photodiode, we sweep $I$ and measure the voltage $V(I)$ and output transmission $T(I)$. We fit the expressions:
\begin{equation}
    V(I) = a_4 I^4 + a_3  I^3 + a_2 I^2 + a_1 I
\end{equation}
\begin{equation}
    T_{\rm cross} = A + B \cos \left ( \frac{I V(I)}{P_\pi} \pi + p_0 \right )
    \label{cross}
\end{equation}
where $a_4, a_3, a_2, a_1, A, B, P_\pi, p_0$ are fitting parameters. Here, $P_\pi$ is the total dissipated power required to induce a $\pi$ phase shift, $p_0$ is the static phase difference between the two interferometer arms, and $1/(A-B)$ is the interferometer extinction ratio. We found that a fourth-order polynomial was required to fit the voltage-current relationship of the heaters, which became non-Ohmic at high currents due to self-heating and velocity saturation of the carriers. If we measured $T(I)$ at the bar port, the latter expression would instead be:
\begin{equation}
    T_{\rm bar} = A - B \cos \left ( \frac{I V(I)}{P_\pi} \pi + p_0 \right )
    \label{bar}
\end{equation}
This yields a mapping between the current $I$ and the programmed phase $\theta_1(I)$ of the form:
\begin{equation}
    \theta_1(I) = p_4 I^4 + p_3  I^3 + p_2 I^2 + p_1 I + p_0
\end{equation}

\subsection*{Coherent matrix multiplication unit (CMXU)}
\begin{figure*}[h]
    \includegraphics[width=7in]{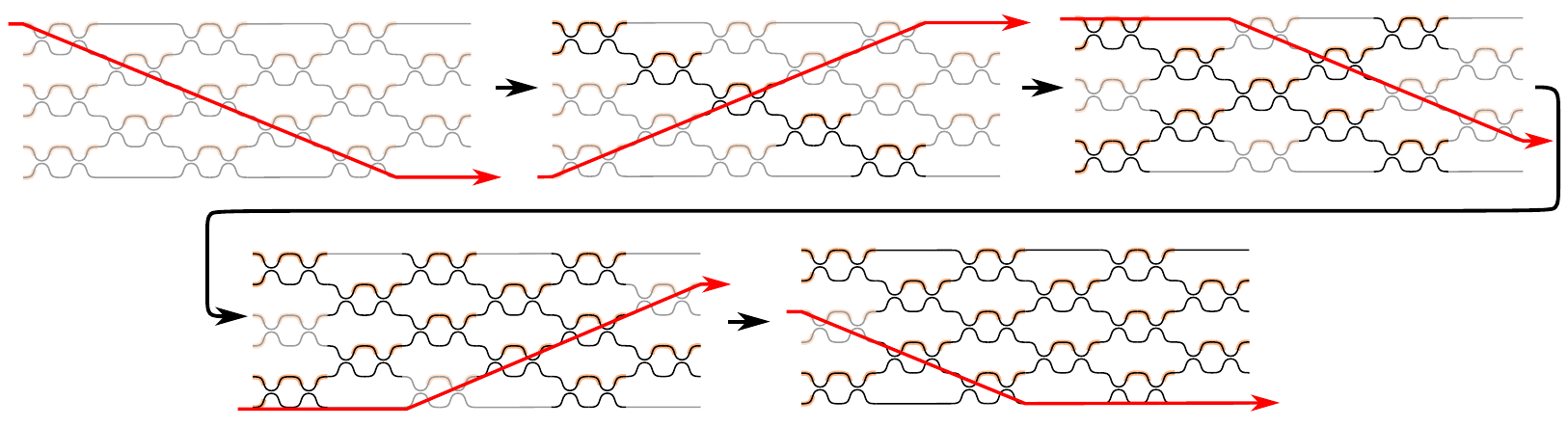}
    \caption{Calibration procedure for internal phase shifters in the CMXU. The devices along the main diagonal and antidiagonal are calibrated first. Once these devices are characterized, the remainder of the phase shifters can be calibrated by programming devices along the main diagonal.}
    \label{characterize_mzi}
\end{figure*}

The procedure for characterizing the CMXU is sketched in Figure \ref{characterize_mzi}. We use the photodiodes at the output of each matrix processor; for the first two layers, we use the photodiode at each nonlinear optical function unit (NOFU), while the last layer is calibrated with the system's receiver.

In an uncalibrated mesh, light will scatter randomly through the circuit as $p_0$ is random for each device. To characterize the circuit, we first input light into the top input (input 1) of the mesh and measure the transmission at the bottom output (output 6). We then optimize the internal phase shifters along the main diagonal in a round robin fashion to maximize the signal at output 6. This procedure deterministically initializes the main diagonal to the cross state ($\theta_1 = 0$), as there is only one possible path between input 1 and output 6. Having initialized the diagonal, we can then calibrate each device along it by sweeping the phase shifter $\theta_1$, measuring $T$ at output 6, and fitting equation \ref{cross} to the data. The antidiagonal, connecting input 6 to output 1, is calibrated in the same way. 

Having characterized the main diagonals, the remainder of the devices can be calibrated in a similar fashion. For instance, inputting light into mode 1 and setting the top left MZI in the circuit, which is already calibrated, to the bar state provides access to the first subdiagonal. The uncalibrated devices can then be characterized with the same procedure as was used for the main diagonal. We show the full calibration sequence in Figure \ref{characterize_mzi}.

\begin{figure*}[h]
    \includegraphics[width=7in]{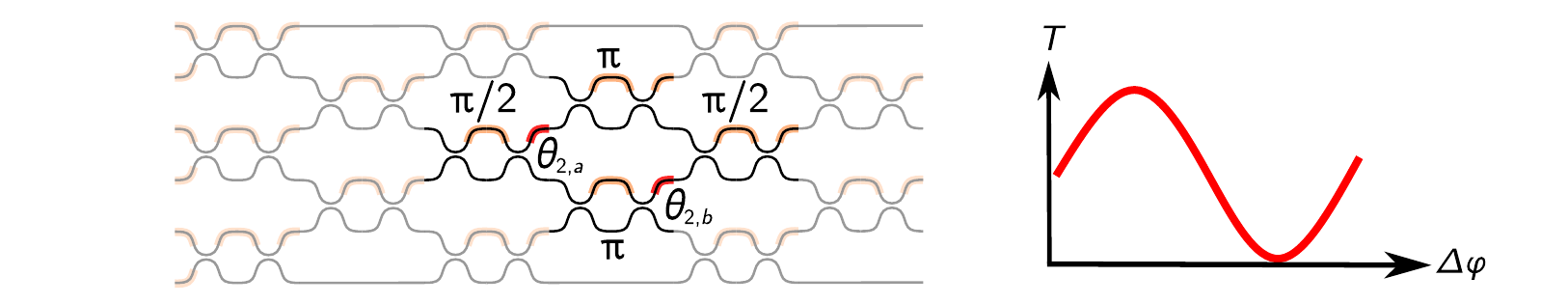}
    \caption{``Meta-MZI'' for calibrating external phase shifters. Two phase shifters in columns $i-1, i+1$ are set to implement a 50-50 beamsplitter. The output transmission of this meta-interferometer, which functions exactly like a discrete MZI, is dependent on the phase difference between the external phase shifters $\Delta \phi = \theta_{2,b}-\theta_{2,a}$.}
    \label{ext_mzi}
\end{figure*}

This protocol calibrates all internal phase shifters $\theta_1$ in a matrix processor. The external phase shifters $\theta_2$ are calibrated using ``meta-MZIs,'' as shown in Figure \ref{ext_mzi}. A ``meta-MZI'' consists of two MZIs in columns $i-1, i+1$ that are programmed to implement a 50-50 beamsplitter ($\theta_1 = \pi/2$). This subcircuit now functions as an effective MZI, where the relative phase difference between two external phase shifters $\theta_{2,a}, \theta_{2,b}$ is equivalent to the setting of the internal phase shifter in a discrete device.

We fix one of the two external phase shifters to $I=0$, sweep the current programmed into the other, and measure the output transmission $T$. Fitting the data to equations \ref{cross}, \ref{bar}, depending on the port $T$ is measured out of, calibrates the static phase difference  $\Delta \phi(I=0) = \theta_{2,b}(I=0)-\theta_{2,a}(I=0)$. Repeating this procedure for all devices produces a linear system of equations that can be inverted to find the static phase $p_0$ for each external heater. More details on this procedure can be found in \cite{prabhu_accelerating_2020}.

\section{Nonlinear optical function unit}
\begin{figure*}
    \includegraphics[width=7in]{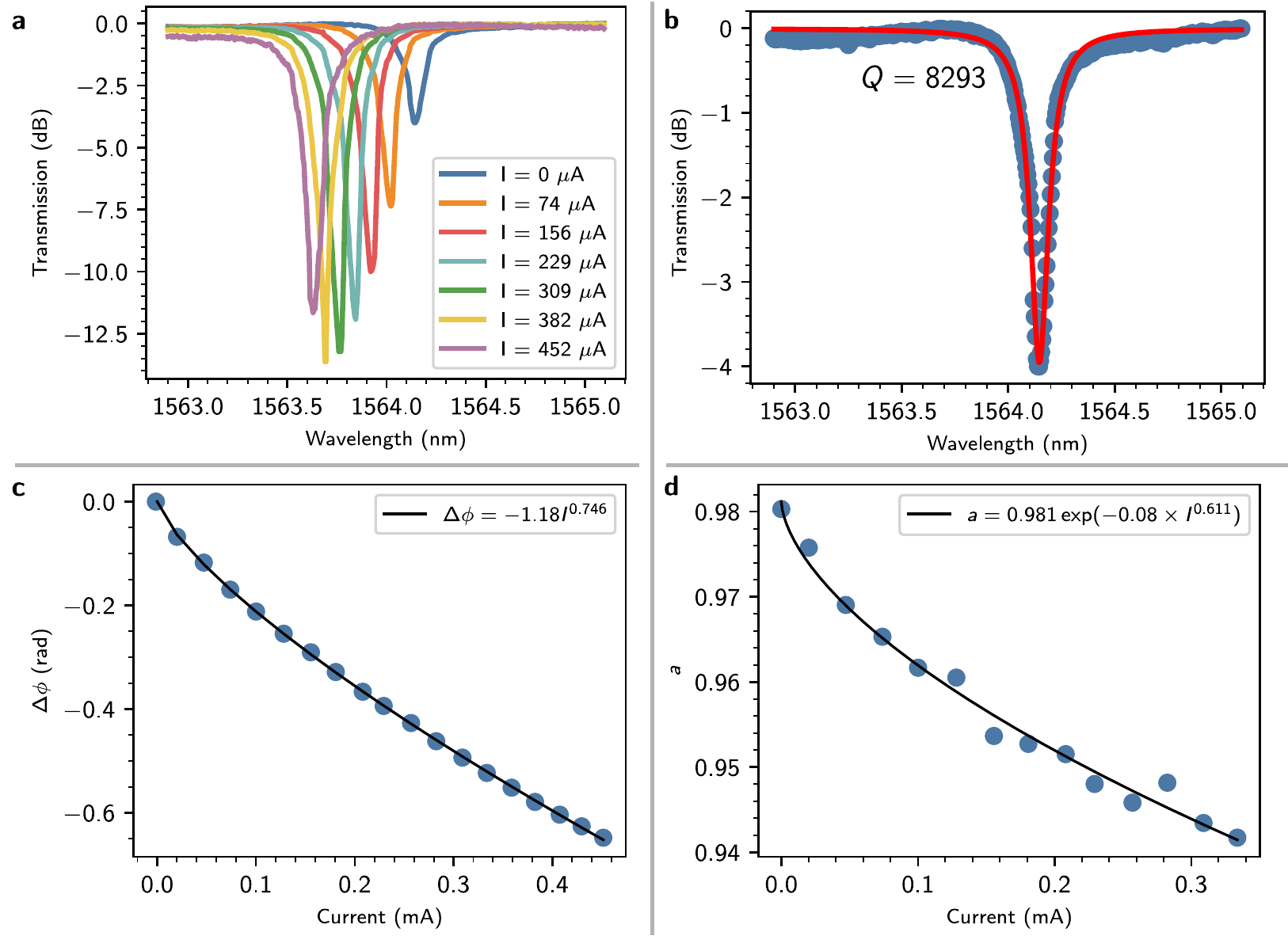}
    \caption{a) NOFU response vs. injected photocurrent. We assume a photodiode responsivity of 1 A/W. The resonator is initially overcoupled, and transitions to undercoupling as the incident optical power is increased. b) NOFU resonance when no power is incident on the photodiode. We measure $Q = 8293$, which limits the photon lifetime to 6.6 ps. c) Phase shift $\Delta \phi$ in cavity vs. incident photocurrent. d) Round-trip amplitude loss $a$ as a function of incident photocurrent. As photocurrent increases more carriers are injected into the waveguide, increasing the loss of the optical signal inside the resonator.}
    \label{NOFU}
\end{figure*}

Figure \ref{NOFU}a shows the measured response of a typical nonlinear optical function unit. The resonator is designed to be overcoupled, as the injection of carriers increases the round-trip loss in the ring. Thus, as current is increased, the resonator transitions through the critical coupling regime to being undercoupled at large incident powers. 

In Figure \ref{NOFU}b, we show the device resonance when no current is injected into the device. The device has a quality factor $Q \approx 8300$, which corresponds to a photon lifetime of 6.6 ps and ensures that the cavity response does not limit the speed of the device. 

The phase response and round-trip attenuation $a$ as a function of the photocurrent $I$ are shown in Figure \ref{NOFU}c and d, respectively. Assuming a photodiode responsivity of $\sim 1$ A/W, we find that about 75 $\mu$W is sufficient to detune the NOFU by a linewidth. As we bias the device to 0.8 V in our experiment, the power consumption during operation is therefore $\sim 60$ $\mu$W.

\section{Correcting hardware errors}
Static component error, such as errors in beamsplitters or transmission losses, affect the accuracy of matrices programmed into the CMXU. Since we use thermal phase shifters for programming the matrix processors, thermal crosstalk between devices will also impact the performance of the device. In this section, we outline the hardware error correction procedures used to obtain high matrix accuracies in the FICONN.

\subsection*{Transmitter correction}

For the transmitter, we corrected thermal crosstalk between components by directly measuring the $12 \times 6$ crosstalk matrix $M$, where $M_\text{ij}$ denotes the crosstalk on channel $i$ produced by an aggressor channel $j$. This quantity was measured by driving channel $i$ and measuring the output transmission $T$ at different current settings for channel $j$. For each measurement, we fit equation \ref{cross} to the data to extract the static phase $p_0$. Thermal crosstalk will cause $p_0$ to vary as a function of the settings in channel $j$ due to parasitic heating; we fit a linear expression to this data to extract the crosstalk coefficient $M_\text{ij}$.

\begin{figure*}
    \includegraphics[width=7in]{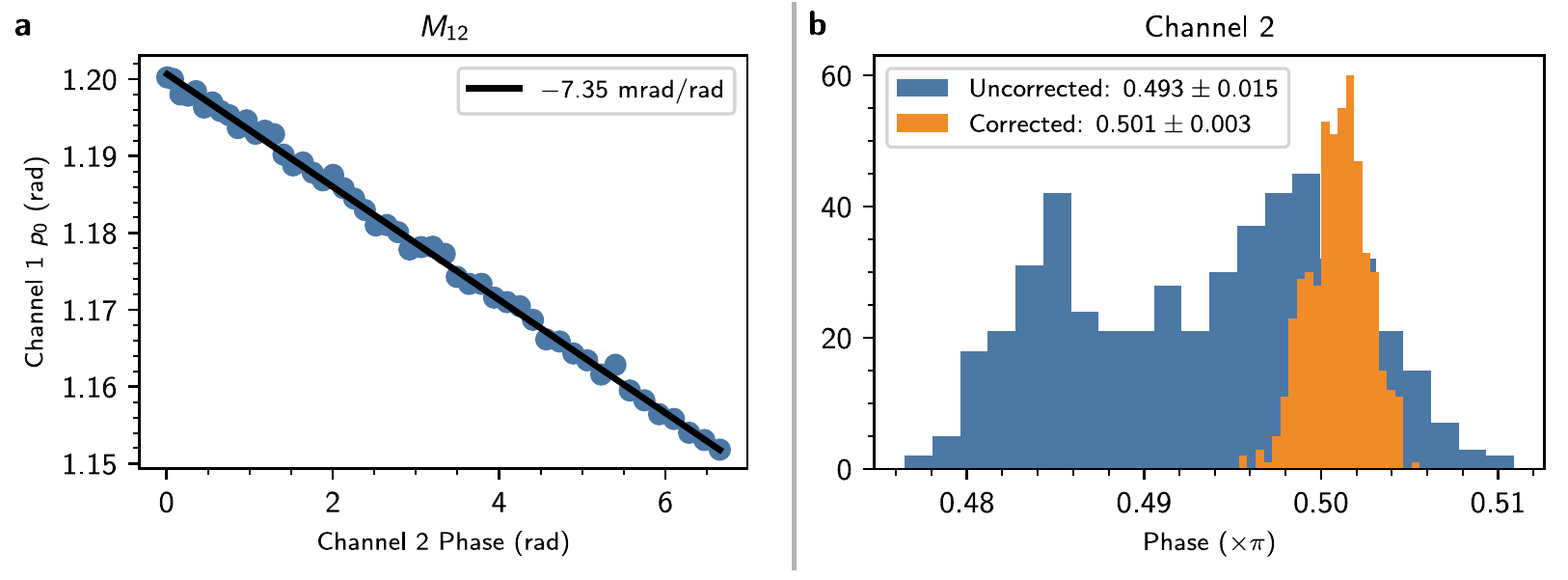}
    \caption{a) To determine the elements of the thermal crosstalk matrix $M$, we drive an aggressor channel $j$ while characterizing the static phase $p_0$ of channel $i$. As an example, here we characterize $M_{12}$ by plotting the static phase of channel 1 as a function of the phase setting of channel 2. We fit a linear function to this data to find a crosstalk coefficient of $M_{12} = -0.00735$. b) We benchmark the effectiveness of thermal crosstalk correction by repeatedly trying to program a channel to $\theta_1 = \pi/2$, while setting all other channels to random values. We then determine the actual phase implemented by measuring the output transmission $T$ and computing $2\arccos \sqrt{T}$. As an example, here we show the results for channel 2, where over 500 random experiments thermal crosstalk correction greatly improves the repeatability of programming a channel to a desired phase.}
    \label{thermal_crosstalk}
\end{figure*}

Figure \ref{thermal_crosstalk}a shows an example of this procedure, where we extracted the crosstalk on channel 1 produced by channel 2. Having obtained $M$, we can now obtain the phase settings $\mathbf{\Phi}$ for a desired programming $\mathbf{\Phi}^\prime$ by computing:
\begin{equation}
    \mathbf{\Phi} = M^{-1}(\mathbf{\Phi}^\prime - \mathbf{\Phi}_0) + \mathbf{\Phi}_0
\end{equation}
where $\mathbf{\Phi}_0$ is the static phase for each channel.
We neglected crosstalk on the external phase shifters of the transmitter, which program the phase of the input $\mathbf{a}^{(1)}$, as we did not have coherent detection directly at the transmitter output. 

In order to benchmark the correction protocol for each transmitter channel we repeatedly attempted to program $\theta_1 = \pi/2$ while setting all other channels to a random phase setting. Shown in Figure \ref{thermal_crosstalk}b is the phase setting actually implemented by the transmitter channel for 500 such experiments, which we extracted by measuring the transmission $T$ and computing $2 \arccos \sqrt{T}$. Thermal crosstalk correction greatly improves both the accuracy and repeatability of each channel; for example, the measured phase on channel 2 improves from $0.493 \pm 0.015$ to $0.501 \pm 0.003$ following correction.

\subsection*{CMXU correction}
Correcting for thermal crosstalk in the CMXU is more challenging. As the CMXU is a mesh of interferometers, changing the programming of aggressor channels can introduce phases and redirect light through the circuit in unexpected ways. These effects are challenging to disentangle from pure thermal crosstalk when the circuit also has other component errors, such as beamsplitter imperfections and device loss, making it difficult to directly measure $M$.

To address this, we instead developed a ``digital twin'' of the hardware, which modeled in software the response of a device with known beamsplitter errors, waveguide losses, and thermal crosstalk. As the effects of all of these imperfections are known \textit{a priori} for Mach-Zehnder interferometer meshes \cite{Bandyopadhyay:21, hamerly_stability_2021}, we can fit a software model, where these imperfections are initially unknown model parameters, to data taken on the real device. If the software model can accurately reproduce measurements from the hardware, the parameters found to describe the device imperfections can be used to deterministically correct errors on the real hardware. We note here that our approach is not a ``black-box'' or neural network model of the device. Our model is based on the physics of how Mach-Zehnder interferometer meshes behave, and thus the parameters we find are realistic and correspond to true physical attributes of the device, such as the error for a particular directional coupler. 

We fit the model to a dataset obtained by programming 300 random unitary matrices into the chip and measuring the response to 100 randomly selected input vectors. Our software model, which is written in \verb|JAX| for auto-differentiability, is fit to the measured data using the limited-memory Broyden–Fletcher–Goldfarb–Shanno (L-BFGS) algorithm. We found that our software model was able to predict hardware outputs with an average fidelity $F = \mathrm{Tr}[U^\dagger_\text{measured} U_\text{software}]/N$ of 0.969 $\pm$ 0.023. The data shown in Figure 2d of the main text was taken by first implementing a ``direct'' programming, where the phase shifter settings were obtained using the Clements decomposition, and then by using the settings obtained from the software model.

\section{Digital model for vowel classification}
We trained a digital model on the vowel classification task to benchmark the performance of \textit{in situ} training on our system. The two models had the same number of neurons ($3 \times 6^2 = 108$), but the weights of the digital model, unlike those of our system, were unconstrained and could be arbitrary real matrices. We trained the system with a tanh nonlinearity, as we obtained very poor performance on the test set using a ReLU function. When training, we normalized the output with a softmax function and used the categorical cross-entropy loss function, as we did in the \textit{in situ} training experiment.

\begin{figure*}
    \includegraphics[width=7in]{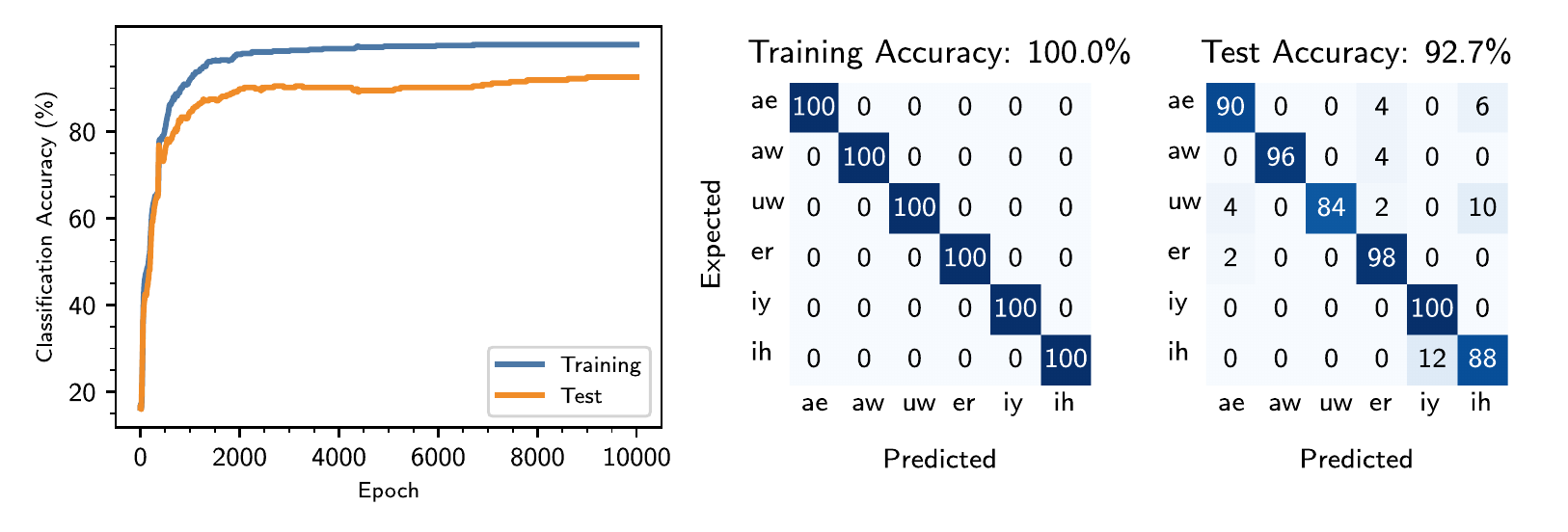}
    \caption{Performance of the digital model on the vowel classification task. The model overfits the training set, achieving 100\% accuracy, but performance on the test set is comparable to the accuracy achieved by our system (92.7\% on the digital model vs. 92.7\% on the FICONN).}
    \label{digital_model}
\end{figure*}

The performance of the digital model is shown in Figure \ref{digital_model}. The performance on the test set is similar to that obtained by our system. However, the digital model is significantly overfit, achieving perfect (100\%) accuracy on the training set. One possible explanation for why our system does not overfit as much is the presence of analog noise, which has been suggested to  function as regularization during DNN training \cite{Camuto_Explicit_2020}. 

\section{Stochastic optimization performs gradient descent on average}

Unlike other derivative-free optimizers, the advantage of the stochastic optimization approach we use is that it performs gradient descent on average. Here we illustrate this by adapting the proof provided in \cite{Cauwenberghs_1992}.

Suppose we are optimizing a DNN with model parameters $\mathbf{\Theta}$ and error functional $\mathcal{L}(\mathbf{\Theta})$. Gradient descent iteratively optimizes the parameters with the update rule:
\begin{equation}
  \Delta \mathbf{\Theta} = -\eta \frac{\partial \mathcal{L}}{\partial \mathbf{\Theta}}
\end{equation}
Assuming $\eta > 0$ and is sufficiently small, this update rule will converge to a local minimum of $\mathcal{L}(\mathbf{\Theta})$. Finite difference methods for analog hardware attempt to compute $\partial \mathcal{L}/\partial \mathbf{\Theta}$ by perturbing one parameter at a time in the system. Each epoch therefore requires $2N$ evaluations of the model on the training set for $N$ model parameters.

Alternatively, one could perturb all model parameters at once by a random vector $\mathbf{\Pi} = [\pi_1, \pi_2, ..., \pi_N]$, where the elements of $\mathbf{\Pi}$ are randomly and independently chosen from an $N$-dimensional hypercube. The update rule here is:
\begin{align}
  \Delta \mathbf{\Theta} &= -\mu \frac{\mathcal{L}(\mathbf{\Theta} + \mathbf{\Pi}) - \mathcal{L}(\mathbf{\Theta} - \mathbf{\Pi})}{2 ||\mathbf{\Pi}||} \mathbf{\Pi}\\
  &= -\frac{\mu}{2 |\pi| \sqrt{N}} [\mathcal{L}(\mathbf{\Theta} + \mathbf{\Pi}) - \mathcal{L}(\mathbf{\Theta} - \mathbf{\Pi})] \mathbf{\Pi}
\end{align}
where $\mu$ is the learning rate. Assuming that the elements of $\mathbf{\Pi}$ are independently drawn from a Bernoulli distribution as $\pm \pi$, we can substitute $||\mathbf{\Pi}||$ as $|\pi| \sqrt{N}$. We note here that $\mu \neq \eta$, and in practice $\mu$ can be much larger than $\eta$ while preserving stable convergence.

We Taylor expand the expression $[\mathcal{L}(\mathbf{\Theta} + \mathbf{\Pi}) - \mathcal{L}(\mathbf{\Theta} - \mathbf{\Pi})]$ as:
\begin{equation}
2\sum_i \frac{\partial \mathcal{L}}{\partial \theta_i} \pi_i
\end{equation}
Substituting this into the update rule, we get:
\begin{align}
  \Delta \mathbf{\Theta} &= -\frac{\mu}{|\pi| \sqrt{N}} \left ( \sum_i \frac{\partial \mathcal{L}}{\partial \theta_i} \pi_i  \right ) \mathbf{\Pi}\\
  &= -\frac{\mu}{|\pi| \sqrt{N}} \left ( \sum_i \frac{\partial \mathcal{L}}{\partial \theta_i} \pi_i  \right ) [\pi_1, \pi_2, ..., \pi_N]
\end{align}
Since the $\pi_i$ are independently chosen, $\mathrm{E}[\pi_i \pi_j] = 0$ if $i \neq j$. Therefore, the expected parameter update $\mathrm{E}[\Delta \mathbf{\Theta}]$ is:
\begin{align}
  \mathrm{E}[\Delta \mathbf{\Theta}] &= -\frac{\mu}{|\pi| \sqrt{N}} \left ( \sum_i \frac{\partial \mathcal{L}}{\partial \theta_i} \mathrm{E}[\pi_i^2] \hat{x}_i \right )\\
  &=  -\frac{\mu}{|\pi| \sqrt{N}} \left ( \sum_i \frac{\partial \mathcal{L}}{\partial \theta_i} |\pi|^2 \hat{x}_i \right )\\
  &= -\frac{\mu |\pi|}{\sqrt{N}} \frac{\partial \mathcal{L}}{\partial \mathbf{\Theta}}
\end{align}
We therefore find that, on average, this procedure performs gradient descent with an effective learning rate $\eta = \mu |\pi| / \sqrt{N}$.

\section{Latency and energy efficiency}

\subsection*{Latency}

The optical propagation delay of the FICONN is the time-of-flight through the photonic circuit, which is $3\tau_\text{CMXU} + 2\tau_\text{NOFU} + \tau_\text{TX to U1} + \tau_\text{U3 to RX} + \tau_\text{U-turn}$. For each subsystem, we computed the waveguide length from the PIC design. While in the actual device waveguides transition between fully-etched and partially-etched geometries, which have different group indices, when estimating latency we conservatively assumed that all waveguides have a ridge geometry, which has a higher group index ($n_g \approx 4.2$) and therefore longer propagation time. The NOFU response time $\tau_{\text{NOFU}}$ includes the cavity lifetime (6.6 ps), as well as time-of-flight through the tunable MZI and delay line. 

We also considered propagation time between the transmitter and first CMXU $\tau_\text{TX to U1}$ and between the last CMXU and the receiver $\tau_\text{TX to U1}$. Due to chip area constraints, we connected different subsystems with U-turns, as shown in Figure 2a of the main text; these waveguides account for an additional 7.5 mm of propagation. In total, the waveguide length from transmitter to receiver is 29.8 mm, which together with the cavity lifetime (6.6 ps $\times$ 2), corresponds to about 435 ps propagation time.

\begin{table}[tb]
\begin{tabular}{l|c|c}
Parameter & Value & Reference \\
\hline
Digital-to-analog conversion (transmitter, 1 GHz) & 26 mW & \cite{sedighi_low-power_dac_2012}\\
Digital-to-analog conversion (transmitter, 50 GHz) & 560 mW & \cite{greshishchev_60_2019}\\
Digital-to-analog conversion (weights) & 27.5 $\mu$W & \cite{ltc1662}\\
Resonant modulator (transmitter) & ~~0.9 fJ/bit at 25 Gb/s $\rightarrow$ 22.5 $\mu$W~~ & \cite{timurdogan_ultralow_2014}\\
Phase shifter (weights, thermal) & 37.5 mW & Measured\\
Phase shifter (weights, MEMS) & 75 $\mu$W & \cite{gyger_reconfigurable_2021}\\
NOFU (injection mode) & 60 $\mu$W & Measured\\
NOFU (depletion mode) & ~~18 fJ/clock cycle~~ & ~~Estimated, see discussion~~\\
Transimpedance amplifier (receiver, 1 GHz) & 57 mW & \cite{sedighi_low-power_tia_2012}\\
Transimpedance amplifier (receiver, 50 GHz) & 313 mW & \cite{ahmed_34gbaud_2018}\\
Analog-to-digital conversion (receiver, 1 GHz) & 2.55 mW & \cite{oh_8b_2020}\\
Analog-to-digital conversion (receiver, 50 GHz) & 150 mW & \cite{adcs}\\
\end{tabular}
\caption{Parameters used for energy efficiency calculation.}
\label{energy}
\end{table}

\subsection*{Energy efficiency}
\noindent The FICONN architecture with $N$ modes and $M$ layers performs $2 M N^2 + 2 (M-1) N$ operations per inference, where the first term accounts for linear matrix operations and the second term refers to the nonlinear activation function. For large $N$ the first term dominates and we approximate the total number of operations as $2 M N^2$. 

The total energy consumption per operation of the system for a single inference can therefore be approximated as $\tau_\text{latency} P_\text{total} / (2 M N^2)$, where $P_\text{total}$ is the total power consumption of the photonics, drivers, and readout electronics and $\tau_\text{latency}$ is the time required for a single inference. The system requires $MN^2$ phase shifters, $N$ transmitters, $N$ receivers, and $MN$ nonlinear optical function units, making the total power consumption $P_\text{total} = MN^2 P_\text{PS} + MN P_\text{NOFU} + N(P_\text{TX} + P_\text{ICR})$. Substituting this into the expression for energy consumption per operation produces equation 2 in the main text. 

Our device performs $2 M N^2 + 2 (M-1) N = 240$ operations per inference, where $M = 3$ and $N = 6$. The phase shifters require about 25 mW per $\pi$ phase shift; as the internal phase shifters only require up to $\pi$ phase shift, while the external phase shifters require up to $2 \pi$, we assume the average power consumption per phase shifter is 37.5 mW. The phase shifter contribution to the energy per operation is therefore $144 \times (37.5~\text{mW}) \times (435~\text{ps}) / (240~\text{OPs}) = 9.8$ pJ/OP, where we include phase shifters for both model parameters and the transmitter. This energy requirement would reduce substantially with the use of undercut thermal phase shifters \cite{dong_thermally_2010}, which reduce power dissipation by an order of magnitude, or MEMS-actuated devices \cite{baghdadi_dual_2021, gyger_reconfigurable_2021}, both of which are available in silicon photonic foundries. The nonlinear optical function unit consumes 60 $\mu$W of power, which contributes about $12 \times (60~\mu\text{W}) \times (435~\text{ps}) / (240~\text{OPs}) = 1.3$ fJ/OP to this total.

We used benchtop electronics in our demonstration to control the devices on chip. In practice, the devices would be controlled by a custom CMOS integrated circuit. To estimate the energy contribution of the electronic driver and readout circuitry, we use values reported in the literature for DACs, TIAs, and ADCs, which are shown in Table \ref{energy}. We assume a clock speed of 1 GHz, as the NOFU in injection mode would have a response time of $\sim$1 ns. The transmitter would require high-speed DACs for programming input vectors into the system, while the receiver would require high-speed TIAs and ADCs for readout. The model parameters can be controlled by lower-speed electronics, as they are not anticipated to change frequently; even when training the system \textit{in situ}, the parameters will only be updated after an entire batch is evaluated. Using these values, we find the worst-case energy consumption of electronics for our device would be $[12 \times (26~\text{mW} + 57~\text{mW} + 2.55~\text{mW}) + 132 \times (27.5~\mu\text{W})] \times (435~\text{ps}) / (240~\text{OPs}) = 1.9$ pJ/OP, yielding a total system performance of 11.7 pJ/OP.

\subsection*{Scaling}
The latency and energy efficiency of the FICONN would be further improved by performing inference on large batches of vectors, as they can be transmitted into the PIC at a faster rate than the end-to-end propagation delay. For $W$ input vectors, the total latency of the system is $\tau_\text{latency} + (W-1)/f_\text{BW}$, where $f_\text{BW}$ is the total bandwidth of the system. For large $W$, the latter term dominates the latency. The system then performs $2f_\text{BW}(MN^2 + (M-1)N)$ TOPS and the energy efficiency becomes:
\begin{equation}
E_\text{OP} \approx \frac{1}{2f_\text{BW}} \left [P_\text{PS} + \frac{P_\text{NOFU}}{N} + \frac{P_\text{TX} + P_\text{ICR}}{MN} \right ]
\end{equation}

The ultimate speed and energy efficiency of our architecture is therefore determined by the rate at which input vectors can be transmitted into the DNN, rather than the end-to-end propagation time. In our implementation the NOFU is realized with a carrier injection device, which would usually be limited to about $\sim$1~ns response time due to the relatively long recombination time in silicon.
While single ns latency is already far lower than what is achievable in conventional electronics, optimizing the NOFU to operate in depletion mode, where the bandwidth is $RC$-limited, could improve this even further to system bandwidths on the order of 50 GHz \cite{sun_128_2019}. Resonant, high-speed modulators in silicon photonics have previously been shown to consume less than 1 fJ/bit \cite{timurdogan_ultralow_2014}. In Table 1 of the main text, the predicted energy consumption of future iterations of the FICONN assumes a clock speed of 50 GHz with the NOFU operating in depletion mode.

To estimate the energy of the NOFU in depletion mode, which would be required to operate at faster clock rates, we assume a capacitance of 200 fF based on reported capacitances for photodiodes ($\sim$15 fF) \cite{novack_germanium_2013}, modulators ($\sim$60 fF, estimated using the depletion capacitance of a 20 $\mu$m microring modulator), and interconnects (200 aF/$\mu$m) \cite{miller_attojoule_2017}. Assuming a 0.3 V drive voltage, this yields an energy consumption of $CV^2 = 9$ fJ/NLOP.

\begin{figure*}
    \includegraphics[width=3.5in]{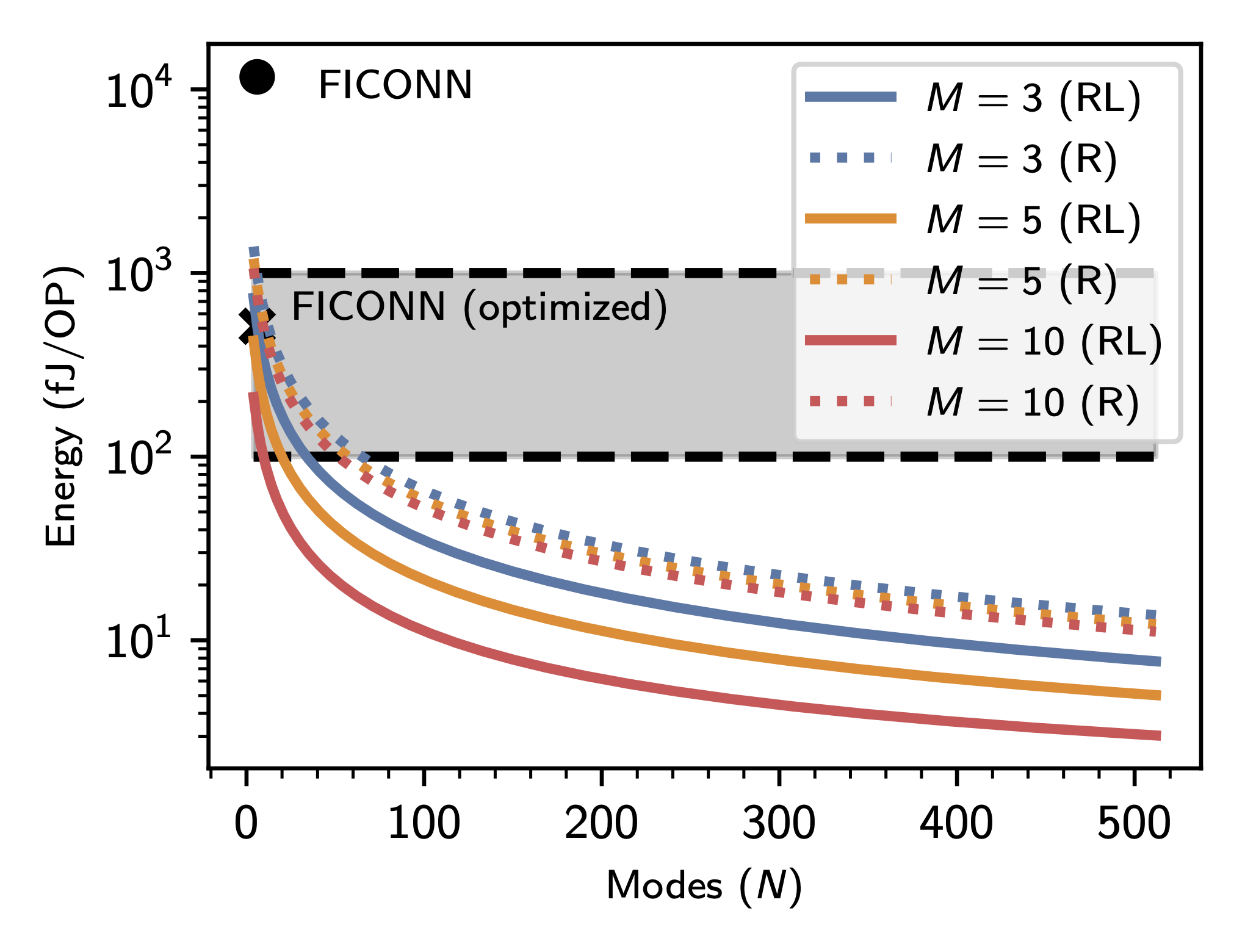}
    \caption{The energy efficiency per operation of a photonic DNN as a function of number of modes ($N$), number of layers ($M$), and whether the architecture is receiverless (RL) or requires readout between each intermediate layer (R). The shaded region indicates the energy efficiencies of current-day electronic ASICs for DNNs ($\sim$0.1$-$1 pJ/OP). An optimized version of our system, with low-power phase shifters and high-speed electronics, would already be competitive with current ASICs.}
    \label{scaling}
\end{figure*}

Performance improves even further for larger system sizes, as the energy cost of the electronics and nonlinearity is amortized by the number of modes $N$. Figure \ref{scaling} shows the expected energy consumption of a receiverless photonic DNN, like our system, and one that reads out optical signals between layers as a function of $N$ and $M$. We show also the current performance region of application-specific integrated circuits for DNNs, which consume approximately 0.1$-$1 pJ per operation \cite{Jouppi2017, shao_simba_2019}, the estimated performance of our current system (11.7 pJ/OP), and that of an optimized version with low-power phase shifters and high-speed electronics (513 fJ/OP). 
For a receiverless, three layer system such as ours, a circuit with $N=34$ modes is sufficient to achieve energy efficiencies below 100 fJ/OP, while $N>380$ attains efficiencies better than 10 fJ/OP. This scaling also improves greatly with the number of layers $M$; for a device with $M=10$ DNN layers, a circuit as small as $N=10$ modes yields efficiencies superior to digital electronics.

We assume a clock rate of 50 GHz in Figure \ref{scaling} to minimize the system latency; however, as the power dissipation of electronics scales nonlinearly with the bandwidth, it may be advantageous to operate at slower clock rates when energy efficiency is the primary concern. Direct processing of optical data would further lower energy consumption, as high-speed DACs \cite{greshishchev_60_2019} are no longer required to encode data onto an optical carrier. High speed readout and digitization of the outputs \cite{ahmed_34gbaud_2018, adcs}, however, is still required. This accounts for the comparatively poorer scaling of energy efficiency for devices that read out between layers. A three layer system that performs intermediate readout needs to have nearly twice as many modes as a receiverless circuit to achieve energy efficiency below 100 fJ/OP. Figure \ref{scaling} shows how the repeated cost of readout results in little improvement in energy efficiency as $M$ increases. As an example, a receiverless system with $M=10, N=114$ attains efficiencies of 10 fJ/OP, while a system with intermediate readout would require $N \approx 575$ to achieve the same efficiency.

\bibliographystyle{naturemag}
\bibliography{Bibliography}